\documentclass[12pt]{article}
\usepackage[margin=25mm]{geometry}

\usepackage{amsthm, amsmath}

\usepackage{times}
\usepackage{bm}
\usepackage{natbib}

\usepackage[plain,noend]{algorithm2e}

\usepackage{amssymb}
\usepackage{mathtools}
\usepackage{subcaption}
\usepackage{afterpage}
\usepackage{booktabs}
\usepackage[tableposition=top]{caption}
\usepackage{multirow}
\usepackage{floatrow}
\usepackage{setspace}
\usepackage{xr-hyper}
\usepackage{xr}
\usepackage{comment}
\usepackage{authblk}

\usepackage{tikz}
\usetikzlibrary{arrows,shapes.arrows,shapes.geometric,shapes.multipart,backgrounds,decorations.pathmorphing,positioning,fit,automata}
\tikzset{
	>=stealth',
	true/.style={
		rectangle,
		draw=black, very thick,
		text width=6.5em,
		minimum height=2em,
		text centered,
		fill=gray, opacity = 0.5},
	punkt/.style={
		rectangle,
		rounded corners,
		draw=black, very thick,
		text width=6.5em,
		minimum height=2em,
		text centered},
	est/.style={
		circle,
		draw=black, very thick,
		text centered},
	shade/.style={
		circle,
		draw=black, very thick, fill=gray!50,
		text centered},
	weight/.style={
		circle,
		draw=black, very thick,
		text width=6.5em,
		minimum height=2em,
		text centered},
	pil/.style={
		->,
		thick,
		shorten <=2pt,
		shorten >=2pt,},
	double/.style={
		<->,
		thick,
		shorten <=2pt,
		shorten >=2pt,},
	dash/.style={
		dashed,
		thick,
		shorten <=2pt,
		shorten >=2pt,},
	dashdouble/.style={
		<->,
		dashed,
		thick,
		shorten <=2pt,
		shorten >=2pt,}
}

\graphicspath{{./art/}}

\DeclareMathOperator*{\argmin}{arg\,min}
\DeclareMathOperator{\vect}{vec}
\makeatletter
\renewcommand{\algocf@captiontext}[2]{\quad #1\algocf@typo. \AlCapFnt{}#2} 
\def\@algocf@capt@plain{top}
\renewcommand{\algocf@makecaption}[2]{%
  \addtolength{\hsize}{\algomargin}%
  \sbox\@tempboxa{\algocf@captiontext{#1}{#2}}%
  \ifdim\wd\@tempboxa >\hsize
    \hskip .5\algomargin%
    \parbox[t]{\hsize}{\algocf@captiontext{#1}{#2}}
  \else%
    \global\@minipagefalse%
    \hbox to\hsize{\box\@tempboxa}
  \fi%
  \addtolength{\hsize}{-\algomargin}%
}
\makeatother

\def\T{{ \mathrm{\scriptscriptstyle T} }}

\def\mI{\mathbb{I}}
\def\p{\partial}
\def\R{\mathbb{R}}
\newcommand{\E}{\mathbb{E}}
\newcommand{\F}{\mathcal{F}}
\newcommand{\de}{\mathrm{d}}

\newcommand{\cov}{\textnormal{cov}}
\newcommand{\obs}{\mathrm{obs}}

\def\nano{\scriptscriptstyle}
\newcommand\hi[1]{^{\nano #1}}
\def\real{\mathbb R}
\newcommand\ca[1]{{\cal{#1}}}
\newcommand\lo[1]{_{\nano #1}}
\newcommand*{\QEDB}{\hfill\ensuremath{\square}}

\newcommand{\BLUE}{\textcolor{black}}
\newcommand{\Blue}{\textcolor{black}}

\newcommand{\blue}{\textcolor{black}}

\makeatletter
\newcommand*{\addFileDependency}[1]{
  \typeout{(#1)}
  \@addtofilelist{#1}
  \IfFileExists{#1}{}{\typeout{No file #1.}}
}
\makeatother

\newtheorem{theorem}{Theorem}
\ifx\example\undefined
\newtheorem{example}[theorem]{Example}
\fi
\ifx\lemma\undefined
\newtheorem{lemma}[theorem]{Lemma}
\fi
\ifx\proposition\undefined
\newtheorem{proposition}[theorem]{Proposition}
\fi
\ifx\remark\undefined
\newtheorem{remark}{Remark}
\fi
\ifx\corollary\undefined
\newtheorem{corollary}{Corollary}
\fi
\ifx\definition\undefined
\newtheorem{definition}[theorem]{Definition}
\fi
\newtheorem{assumption}{Assumption}

\DeclarePairedDelimiter\floor{\lfloor}{\rfloor}

\usepackage{verbatim}
\immediate\write18{texcount -tex -sum  \jobname.tex > \jobname.wordcount.tex}

\providecommand{\keywords}[1]
{
	\small	
	\textbf{\textit{Keywords---}} #1
}
\title{Functional principal component analysis  with informative observation times}
\author{Peijun Sang$^{1}$, Dehan Kong$^{2}$, Shu Yang$^{3}$  \\
	\small $^{1}$Department of Statistics and Actuarial Science, University of Waterloo\\
	\small $^{2}$Department of Statistical Sciences, University of Toronto\\
	\small $^{3}$ Department of Statistics, North Carolina State University
}
\date{} 

\begin{document}

	\maketitle
	
	\begin{abstract}
Functional principal component analysis has been shown to be invaluable for revealing variation modes of longitudinal outcomes, which serves as important building blocks for forecasting and model building. Decades of research have advanced methods for functional principal component analysis often assuming independence between the observation times and longitudinal outcomes. Yet such assumptions are fragile in real-world settings where observation times may be driven by outcome-related reasons. Rather than ignoring the informative observation time process, we explicitly model the observational times by a general counting process dependent on time-varying prognostic factors. Identification of the mean, covariance function, and functional principal components ensues via inverse intensity weighting. We propose using weighted penalized splines for estimation and establish consistency and convergence rates for the weighted estimators. Simulation studies demonstrate that the proposed estimators are substantially more accurate than the existing ones in the presence of a correlation between the observation time process and the longitudinal outcome process. We further examine the finite-sample performance of the proposed method using the Acute Infection and Early Disease Research Program study.
	\end{abstract}
	

\section{Introduction}
Longitudinal data have been extensively studied in the literature
of statistics. Our research is motivated by the investigation
of the disease progression in HIV-positive patients. Highly active
antiretroviral therapy (HAART) has been shown to be an effective treatment
for HIV in improving the immunological function 
and delaying the progression to AIDS \citep{hecht2006}. Our goal
is to study the mean trend and variation mode of CD4 counts, an indicator of immune function,  over time
after treatment initiation, which is of major importance: first, it
depicts a whole picture of how the disease evolves over time and
thus provides new insights into the treatment mechanism. Second, it enables
the prediction of disease progression and helps patients manage the
disease better. Third, such information can also be used to design
optimal treatment regimes for better clinical outcomes \citep{guo2021estimation}. 

Parametric random effect models \citep{laird1982} and generalized
estimating equations models \citep{liang1986} are commonly adopted
to fit longitudinal data; see \citet{diggle2002} for a comprehensive
overview. Though the latter does not need to specify the parametric distribution
of the longitudinal response, it imposes
a specific form of the mean response. 
To better understand the complexity of real-world data, Figure \ref{fig:CD4}(a)
shows the trajectories of CD4 counts at the follow-up visits from
five randomly selected patients in the motivating application, where the number and timing of the
visits differ from one to the next. Extracting useful information
from such data has become a challenging statistical problem.

Functional data analysis offers a nonparametric means to modeling
longitudinal data at irregularly spaced times. Repeated measurements
of a longitudinal response from a subject are regarded as sparsely
sampled from a continuous random function subject to measurement errors.
Moreover, the underlying true random function is typically modeled
in a nonparametric manner, thus avoiding model misspecification suffered
from the two aforementioned approaches. To estimate the mean and covariance
functions of the underlying continuous function from sparse observations,
existing approaches usually assume that the observation times are
independent of the longitudinal responses and then apply nonparametric
smoothing techniques such as kernel smoothing to the aggregated observations
from all subjects; see \citet{yao2005}, \citet{li2010uniform} and
\citet{zhang2016} for instance.

Yet the independence assumption of the observation times and responses
is restrictive in practice; e.g., patients
with deteriorative health conditions may be more likely to visit the health
care facilities \citep{phelan2017illustrating}. Without
addressing the informative observation time process, the study results
can be biased and misleading (\citealp{lin2004analysis}; \citealp{sun2021recurrent}). \Blue{\cite{xu2024bias} considered using a marked point process to model the informative visit times in longitudinal studies. But their work assumes that both the longitudinal outcome process and the latent process used to define the intensity function of the point process are Gaussian, which may not hold in practice.} \BLUE{To address the same issue, \cite{weaver2023functional} assumed that both the intensity function of the point process and the longitudinal outcome process depend on a positive latent factor. This assumption is slightly restrictive and can hardly be verified since it implies the dependence between observation times and the longitudinal outcome can be completely explained through this single latent factor. 
}
In this article, we propose to model the observation time process
by a general counting process with an intensity function depending
on time-dependent confounders. \Blue{But it should be noted that the time-dependent confounders can be just functions of the observed outcome themselves.}
To account for the effect of the observation
time process when estimating the mean function, we leverage the inverse
of the intensity function at each observation time point as its weight
and then apply penalized B-spline functions to the aggregated observations.
This idea is further extended to estimating the covariance function
with the tensor product of B-spline bases, weighted by a product of
the inverse of the intensity functions at the two time points,
to correct the selection bias of the pairs of observations. Variation
modes can thus be visualized through an eigen-decomposition on the
estimated covariance function, which is referred to as functional
principal component analysis.

The proposed functional principal component analysis accounts for the dependence between the response
process and the observation time process via inverse intensity function
weighting. This
fills an important gap in the literature as traditional approaches often assume that response observations
are independent of the observation times, which however is likely
to be violated in real-world studies. Moreover, we establish
consistency and convergence rates of our proposed estimator when estimating
the mean function, covariance function, and functional principal components of a random function. Numerical
studies demonstrate that in contrast to the traditional approaches,
our approach can yield consistent estimates when the response observation
times are indeed correlated with the underlying response process. 

\section{Basic Setup}

\subsection{Functional principal component analysis and observation time process} 

Suppose that $X$ is a random function defined on a compact set $\mI \subset \R$. Let $L^2(\mI)$ denote the collection of measurable square-integrable functions on $\mI$. 
Furthermore, we assume that $\int_{\mI} \E\{X^2(t)\} \de t < \infty$.  Let $\mu(t) = \E\{X(t)\}$ and $C(s, t) = \cov\{X(s), X(t)\}$ denote the mean function and the covariance function of $X$, respectively. Then we can define the covariance operator $C: L^2(\mI) \rightarrow L^2(\mI)$ that satisfies $(Cf)(t) = \int_{\mI} C(s, t)f(s) \de s$ for any $f \in L^2(\mI)$. It follows from Mercer's theorem that there exists an orthonormal basis $(\varphi_j)_j$ of $L^2(\mI)$ and a sequence of nonnegative decreasing eigenvalues $(\kappa_j)_j$ such that
$
C(s, t) = \sum_{j = 1}^{\infty} \kappa_j \varphi_j(s)\varphi_j(t). 
$
The eigenfunctions of $C$, $\varphi_j$'s are also referred to as functional principal components of $X$. 
In fact, $X$ admits the following Karhunen-Lo\`eve expansion, 
$X(t) = \mu(t) + \sum_{j = 1}^{\infty} \zeta_j \varphi_j(t)$, where $\zeta_j = \int_{\mI} \{X(t) - \mu(t)\} \varphi_j(t) \de t$ is called the $j$th functional principal component score of $X$ and satisfies $\E(\zeta_j \zeta_k) = \delta_{jk} \kappa_j$, where $\delta_{jk} = 1$ if $j = k$ and $0$ otherwise. 
The expansion is useful to approximate an infinite-dimensional random function because 
approximating $X(t)$ by $ \mu(t) + \sum_{j = 1}^{p} \zeta_j \varphi_j(t)$ {yields the minimal mean squared error when using  an arbitrary orthonormal system consisting of $p$ functions for any $p \in \mathbb{N}^+$}. 
Additionally, functional principal component analysis enables us to understand variation modes of this random function, as it displays the greatest variations along with the directions of principal components.  

In practice, a fully observed trajectory of a random function may not be accessible due to various practical hurdles and is only observed at sparsely and irregularly spaced time points. 
To describe the irregularly-spaced observation time process for observing $X_i(t)$, let the set of visit times be $0\leq t_{i1}<\ldots<t_{im_{i}} \leq \tau$,
where $m_{i}$ is the total number of observations, and $\tau$ denotes the predetermined study end time. Therefore, the domain of the random function $X(t)$ is $\mI = [0, \tau]$.
In stark contrast
to the regular time setting, the observed time points are allowed to vary from one
subject to another. Let $X_{ij} = X_i(t_{ij}) + \epsilon_{ij}$
denote the noisy observation of the $i$th random function at time $t_{ij}$, where $\epsilon_{ij}$ is the measurement error. Our primary interest  is to perform functional principal component analysis from  observations $\{X_{ij}: j = 1, \ldots, m_i, i = 1, \ldots, n\}$. 
\citet{yao2005} and \citet{li2010uniform} address this problem under the assumption that the observed time points  are independently and identically distributed, and $X\lo{ij}, t\lo{ij}$ and $m_i$ are independent of each other for subject $i$. 
However, in practice, whether or not there exists an observation at one particular time point often depends on the response process. Therefore, analysis of such data requires assumptions on the mechanism for the observation time process.  

\subsection{Informative observation times}

Let $N_{i}(t)$ be the \Blue{general} counting process
for the observation times; that is, $N_{i}(t)=\sum_{j=1}^{\infty}I(t_{ij}\leq t)$ for $t \in [0, \tau]$. 
We use overline to denote the history;
e.g., $\overline{X}_{i}(t) =\{X_{i}(u):0\leq u\leq t\}$ is the
history of the stochastic process $X$ until time $t$ for the $i$th subject.  \Blue{It is possible that the dependence between the longitudinal outcome and the observation times can be explained by merely using functions of the observed outcomes. Next we focus on a more complicated scenario, where an auxiliary process is also involved in inducing the dependence of the outcome and observation times.}
In addition to the response process, we also observe a covariate process $Z_i(t)$ \textcolor{black}{that is related $X_i(t)$ and $N_i(t)$, which can be multivariate, time-independent, or time-varying.}
Let $\text{\ensuremath{\overline{X}}}_{i}^{\obs}(t)=\{X_{i}(s):\de N_{i}(s)=1,0 \leq s \leq t\}$
and $\overline{N}_{i}(t)=\{N_{i}(s):0\leq s\leq t\}$ be the history
of observed variables and observation times through $t$,
respectively.  We
denote the observed history of variables for subject $i$ at time $t$ as {$\overline{O}_{i}(t)=\{\overline{X}_{i}^{\obs}(t-),\overline{N}_{i}(t-), \overline{Z}_{i}^{\obs}(t-)\}$}, where $t-$ indicates the time up to but excluding $t$. 
We use $\mathcal{F}^*_{it}$ to denote the filtration generated by $\overline{O}_i(t)$ and $X_i(t)$, and \blue{$E\{\de N_i(t) \mid \mathcal{F}^*_{it}\} $ denotes the estimated number of observation made in $[t, t + \de t)$, given the observed history up to $t$ for the $i$th subject.}
Let $\lambda\{t\mid\overline{O}_i(t)\}=\E\{\de N_i(t)\mid\overline{O}_i(t)\}/ \de t$ denote the conditional intensity function of $N_i(t)$ given the observed history up to $t$, but not including $t$.
For the above notations, we suppress $i$ to denote their population counterparts. 
In practice, the irregular observation times can be due to a number of reasons that may be related to subjects' responses, in which case, we say that the observation times are informative. 
In this case, ignoring the observation time process leads to biased results for the response variable. 
Similar to the missing data literature, we require a further assumption  to identify the mean and variance functions of $X(t)$ under an informative observation time process. 
\begin{assumption}\label{asmp:idenA}
(i) {$\E\{\de N_i(t)\mid \overline{O}_{i}(t), X_i(t)\}=\E\{\de N_i(t)\mid\overline{O}_i(t)\}$,
and (ii) $\lambda\{t\mid\overline{O}_i(t)\} >0$}
almost surely for $i = 1, \ldots, n$.
\end{assumption}
Assumption \ref{asmp:idenA}(i) implies that the observed history collects all prognostic variables that affect the observation time process. It is plausible when $\overline{O}_i(t)$ includes the past observed responses $\overline{X}^{\obs}(t-)$, historical observation pattern $\overline{N}(t-)$, and past observed important auxiliary confounder process $\overline{Z}^{\obs}(t-)$ that is related to both observation time and response.  
Assumption \ref{asmp:idenA}(ii) suggests that all subjects have a positive probability of visiting at any time $t$. 
Assumption \ref{asmp:idenA} is key toward identification, see $\mathsection$\ref{sec:iden}; however, it is not verifiable based on the observed data and thus requires careful consultation of subject matter knowledge. 

\subsection{Identification via inverse intensity function weighting \label{sec:iden}}
We show that Assumption \ref{asmp:idenA} leads to the identification
of $\mu(t)$, $C(t,s)$, and $\varphi_j(t)$ by providing a brief outline of the proof below, while a detailed proof is deferred to $\mathsection$S.1 in the supplementary material. First, by the law of total expectation,
we have {for any $t \leq \tau$,}
\begin{equation}
\E[X(t)\lambda^{-1}\{t\mid\overline{O}(t)\}\de N(t)]=\E[X(t)\lambda^{-1}\{t\mid\overline{O}(t)\}\E\{\de N(t)\mid\F^{*}_t\}]=\mu(t)\de t.\label{eq-exp}
\end{equation}
Weighting by $\lambda^{-1}\{t\mid\overline{O}(t)\}$ serves to create
a pseudo-population in which the observation time process is no longer
associated with $X(t)$ as if the observed responses were sampled
completely at random. Thus, $\mu(t)$ is identifiable. 

Next, assuming $s<t$ and by the double use of the law of total expectation,
we have 
\begin{eqnarray}
 &  & \E[\{X(t)-\mu(t)\}\{X(s)-\mu(s)\}\lambda^{-1}\{s\mid\overline{O}(s)\}\de N(s)\lambda^{-1}\{t\mid\overline{O}(t)\}\de N(t)]\nonumber \\
 & = & \E[\{X(t)-\mu(t)\}\{X(s)-\mu(s)\}\lambda^{-1}\{s\mid\overline{O}(s)\}\de N(s)\lambda^{-1}\{t\mid\overline{O}(t)\}\E\{\de N(t)\mid\F^{*}_t\}]\nonumber \\
 & = & \E[\{X(t)-\mu(t)\}\{X(s)-\mu(s)\}\lambda^{-1}\{s\mid\overline{O}(s)\}\de N(s)\de t]\nonumber \\
 & = & \E[\{X(t)-\mu(t)\}\{X(s)-\mu(s)\}\lambda^{-1}\{s\mid\overline{O}(s)\}\E\{\de N(s)\mid X(t), \F^{*}_s\}\de t]\nonumber \\
 & = & \E[\{X(t)-\mu(t)\}\{X(s)-\mu(s)\}\de t\de s]=C(t,s)\de t\de s.\label{eq-cov}
\end{eqnarray}
Weighting by $\lambda^{-1}\{s\mid\overline{O}(s)\}\lambda^{-1}\{t\mid\overline{O}(t)\}$
serves to create a pseudo-population in which the observation time
process is no longer associated with $\{X(t)-\mu(t)\}\{X(s)-\mu(s)\}$. Hence, $C(t,s)$ is identifiable.

\section{Estimation}
\label{sec:estimation}

In practice, the intensity function for the observation time process is unknown and requires modeling and estimation.
Following \citet{lin2004analysis} and \cite{yang2018modeling,yang2019bmk}, we assume the intensity function follows a proportional intensity function $\lambda\{t\mid\overline{O}(t)\}=\lambda_{0}(t)\exp[g\{\overline{O}(t)\}^{\T}\beta]$,
where $g(\cdot)$ is a pre-specified multivariate function of $\overline{O}(t)$.
Let $\theta=\{\lambda_{0}(t),\beta\}$. 
Under Assumption \ref{asmp:idenA}, the estimator of $\theta$, denoted by $\widehat{\theta}=\{\widehat{\lambda}_{0}(t),\widehat{\beta}\}$, can be obtained from the standard software.

We treat the estimated intensity function, $\{\widehat{\lambda}_0(t_{ij})\}^{-1}\exp[-g\{\overline{O}_i(t_{ij})\}^{\T}\widehat{\beta}]$, as the sampling weight of $X_{ij}$. To estimate the mean function, because we cannot accurately recover each trajectory of $X_i$ from {sparse and noisy} observations, we propose using weighted penalized splines to borrow information from aggregated observations from all subjects. 
In particular, let $0 = \xi_0  < \xi_1 \leq \cdots \leq \xi_K < \xi_{K + 1} = \tau$ be a sequence of knots. 
The number of interior knots $K = K_n = n^{\eta}$
with $0 < \eta < 0.5$ {being} a positive integer such that $\max_{1 \leq k \leq K + 1} |\xi\lo k - \xi\lo{k - 1} | = O(n^{-\eta})$. 
Let $\ca S\lo n$ be the space of polynomial splines of order $l \geq 1$ consisting of functions $h$ satisfying: (i) in each subinterval, $h$ is a polynomial of degree $l - 1$; and 
(ii) for $l \geq 2$ and $0 \leq l\hi\prime \leq l - 2$, $h$ is $l\hi\prime$
times continuously differentiable on $[0, \tau]$. 
Let  $\{B_j(\cdot), 1 \leq j \leq q_n\}$, $q_n = K_n + l$, be the normalized B-spline basis functions of $\ca S\lo n$. Then for any $h \in \ca S_n$, there exists  $\gamma = (\gamma_1, \ldots, \gamma\lo{q\lo n})^{\T} \in \R^{q_n}$ such that $h(t) = \sum_{j = 1}^{q_n} \gamma_j B_j(t) = \gamma^{\T}B(t)$ for $t \in [0, \tau]$. 
To account for the effect of the observation time process on estimating the mean function, we define the weight 
\begin{equation} \label{eq-meanweight}
w\lo{ij}(\mu) = \{\widehat{\lambda}_0(t_{ij})\}^{-1}\ \exp[-\widehat{\beta}^{\T}g\{\overline{O}_i(t_{ij})\}]
\end{equation} 
for the $j$th observation of the $i$th subject, where $j = 1, \ldots, m_i$ and $i = 1, \ldots, n$. Let $m$ be a positive integer, smaller than $l$. 
Suppose the penalty term in the penalized splines is 
\begin{equation*} 
\int_0^{\tau} \gamma^{\T}\left\{B\hi{(m)}(t)\right\}\hi{\otimes 2} \gamma~\de t,
\end{equation*}
where $a \hi{\otimes 2} = aa^{\T}$ for any matrix or column vector $a$.
Consequently, the penalty matrix is $Q\lo{\mu} = \int_0^{\tau} \left\{B\hi{(m)}(t)\right\}\hi{\otimes 2} ~\de t$.
We then estimate the mean function by $\widehat{\mu}(t) = B^{\T}(t)\widehat{\gamma}\lo{\mu}$ with 
\begin{align}
\widehat{\gamma}\lo{\mu} = \argmin_{\gamma \in \R \hi {q\lo n}} \left[\sum_{i=1}^{n} \sum_{j = 1}^{m_i} \{X_{ij} - B(t_{ij})^{\T}\gamma\}^2 w_{ij}(\mu) + \frac{\lambda_{\mu}}{2} \gamma^{\T}Q\lo{\mu}\gamma \right],
\label{eq-estmean}
\end{align}
where $\lambda_{\mu} > 0$ is a tuning parameter controlling the roughness of the estimated mean function.

Next, we present an estimator of the covariance function. 
Let $G_i(t_{ij}, t_{il}) = \{X_{ij} - \widehat{\mu}(t_{ij})\} \{X_{il} - \widehat{\mu}(t_{il})\}$ be the raw estimate of the covariance function evaluated at $(t_{ij}, t_{il})$. By  \eqref{eq-cov}, we introduce 
\begin{equation} \label{eq-covweight}
w\lo{ijl}(C) = \{\widehat{\lambda}_0(t_{ij}) \widehat{\lambda}_0(t_{il})\}^{-1} \exp\left(-\widehat{\beta}^{\T}[g\{\overline{O}_i(t_{ij})\} + g\{\overline{O}_i(t_{il})\}] \right), 
\end{equation}  
where $j, l = 1, \ldots, m_i$  and $i = 1, \ldots, n$, to account for the effect of the observation time process on estimating the covariance function. We use the tensor product of $B_j(t)$'s to estimate this bivariate covariance function. More specifically, $C(t, s)$ is approximated by $\sum_{1 \leq j_1 \leq j_2 \leq q_n}\eta\lo{j\lo 1 j \lo 2}B\lo{j\lo 1}(t) B\lo {j \lo 2}(s)$. To ensure that $C(t, s) = C(s, t)$, we require $\Xi = (\eta\lo{j\lo 1 j \lo 2})$ to be a $q_n \times q_n$ symmetric matrix. Denote $D(t, s) = B(t) \otimes B(s)$, which is a vector of length $q\lo n \hi 2$, and $\eta = \vect(\Xi)$.
Then the estimated covariance function is 
$\widehat{C}(t, s) = \sum_{j_1,j_2 = 1}^{q\lo n} \widehat{\eta} \lo{j_1j_2} B\lo{j_1}(t)B\lo{j_2}(s)$, 
where $\widehat{\Xi}$ is obtained by solving the minimization problem:
\begin{equation}
\widehat{\Xi} = \argmin_{\Xi = \Xi^{\T}} \left[ \sum_{i=1}^{n} \sum_{1 \leq j \neq l \leq m_i} \left\{G_i(t_{ij}, t_{il})  -  \sum\lo{j_1, j_2 = 1}^{q_n} {\eta} \lo{j_1j_2} B(t\lo{ij})B(t\lo{il})\right\}^2 w_{ijl}(C)  
+ \frac{\lambda\lo C}{2} \eta^{\T} Q\lo C \eta \right]. 
\label{eq-estcov}
\end{equation}
Here $Q\lo C$ is a $q\lo n \hi 2 \times q\lo n \hi 2$ penalty matrix with $(j_1, j_2)$th entry being 
$$
\int_0^{\tau} \int_0^{\tau} \left\{\sum_{i +  j = m}\binom{m}{i} \p\hi {i} D\lo{j\lo 1}(t, s) \p \hi {j} D\lo{j\lo 2}(t, s) \right\} \de t \de s
$$
\citep{lai2013}, and $\lambda_C > 0$ is a tuning parameter that controls the trade-off between fidelity to the data and plausibility of $C(t, s)$. The functional principal components are then estimated by solving 
\begin{equation}
\int_0^{\tau} \widehat{C}(s,t)\varphi_j(s) \de s = \widehat{\kappa}_j \varphi_j(t)
\label{eq-estphi}
\end{equation}
subject to $\int_0^{\tau} \varphi_j^2(t) \de t = 1$ and $\int_0^{\tau} \varphi_j(t) \varphi_k(t) \de t = 0$ when $j \neq k$. 
Detailed steps for solving these equations can be found in Chapter 8.4 of \cite{ramsay2005}. 

In the following numerical implementations, we take $m = 2$. The generalized cross-validation  is used to choose the tuning parameters $\lambda_{\mu}$, $\lambda_C$ and the number of basis function $q_n$. In particular, let $Y$ denote the vector of length $N = \sum_{i = 1}^n m_i$ consisting of observations $X_{ij}, j = 1, \ldots, m_i, i = 1, \ldots n$.  Let $W = \text{diag}\{w_{ij}(\mu), j = 1, \ldots, m_i, i = 1, \ldots, n\}$ and $Y_w = W^{1/2}Y$. According to Chapter 3 of \citet{gu2013smoothing}, the generalized cross-validation score for \eqref{eq-estmean} is 
$$
V_{\mu}(\lambda_{\mu}) = \frac{N^{-1}Y_w^{\T}\{I - A_w(\lambda_{\mu})\}^2 Y_w}{[N^{-1}\text{tr}\{I - A_w(\lambda_{\mu})\}]^2},
$$
where $I$ is an $N \times N$ identity matrix and $A_w(\lambda_{\mu})$ is the so-called smoothing matrix satisfying $\widehat{Y}_w = W^{1/2}\widehat{Y} = A_w(\lambda_{\mu})Y_w$. An explicit form of $A_w(\lambda_{\mu})$ can be found in  (3.12) of \citet{gu2013smoothing}. We select $\lambda_{\mu}$ by minimizing $V_{\mu}(\lambda_{\mu})$. The smoothing parameter $\lambda_C$ for the covariance function estimate defined in  \eqref{eq-estcov} is chosen in a similar manner. Moreover, the number of basis functions $q_n$ is selected by gradually increasing its value in a grid until it leads to a significant decrease in the generalized cross-validation score; see $\mathsection$S.2.2 of the supplementary material for details.

\section{Theoretical Properties}
\subsection{Large sample properties of the mean function estimator}

For $d \in N^+$, let $C \hi d ([0, \tau])$ denote the class of functions with continuous $d$th derivatives over $[0, \tau]$. Without loss of generality, we assume  $\tau = 1$ in the theoretical analysis. {Below, we present the regularity assumptions for deriving the large sample properties for proposed mean and covariance  function estimators, as well as the estimated functional principal components.} 
\begin{assumption} \label{ass: continuous}
	The true mean function of $X$, $\mu(\cdot)$, belongs to $C\hi d ([0, \tau])$ for some $d \geq \max(2, m)$.
\end{assumption}

\begin{assumption} \label{ass: spline}
The knots are equally spaced in $\ca S_n$. The order of the spline functions satisfies $l \geq d$ and $l > m$. 
\end{assumption}
\begin{remark} \label{rmk:approximation}
	Assumptions \ref{ass: continuous} and \ref{ass: spline} ensure that  there exists a spline function $\tilde{\mu}(\cdot) = B^{\T}(\cdot) \tilde{\gamma} \in \ca S_n$ such that
	$
	\| \mu - \tilde{\mu}\|_{\infty} = O(q_n^{-d}). 
	$
The equal-spaced knots assumption is used for the convenience of deriving the decay rate; see Proposition 4.2 of \cite{xiao2019asymptotic}. Proof of this result is similar to that of Lemma 1 of \cite{smith1979efficient} and is omitted here. \blue{Without the equal-spaced knots assumption, deriving the decay rate of the eigenvalues of the relevant penalty matrix would be more challenging. This is left for future research.}

\end{remark}

\begin{assumption} \label{ass: finite4}
Therefore exists some constant $\delta > 2$ such that
$\E(\|X\|_{\infty}^{\delta}) < \infty$. 
\end{assumption}

\begin{assumption} \label{ass: error}
	The random errors $\epsilon_{ij}$'s are independent and identically distributed with mean 0 and $\E(\|\epsilon\|_{\infty}^{\delta}) < \infty$, where $\delta$ is defined in Assumption \ref{ass: finite4} and $\epsilon$ denotes the random process of the error.
\end{assumption}

\Blue{Establishing the uniform convergence rate on the estimated mean function entails strong moment conditions on $X$ and $\epsilon$ as in Assumptions \ref{ass: finite4} and \ref{ass: error}. Similar assumptions are considered in \cite{li2010uniform} and \cite{zhang2016}.}

\begin{assumption} \label{ass: intensity}
	In the intensity function $\lambda(t) = \lambda_0(t)\exp[g\{\overline{O}(t)\}^{\T}\beta_0]$,
	$\lambda_0(t)$ \Blue{belongs to $C^{p}([0, \tau]$) for some $p \geq d$} and is strictly positive, and $g\{\overline{O}(t)\}$ is almost surely bounded over $[0, \tau]$. 
\end{assumption}
This assumption specifies a smoothness property for the baseline intensity to ensure that a desirable convergence rate can be achieved when replacing the true intensity function with the estimated one in  \eqref{eq-meanweight}. This assumption is commonly adopted in a semiparametric Cox model \citep{cox1972} for modeling the intensity function for a counting process. 
We can estimate $\beta$ by the partial likelihood approach, the cumulative baseline intensity function $\Lambda_0(t) = \int_0^t \lambda_0(s) \de s$ by Breslow's estimator, and further $\lambda_0(t)$ by a kernel-smoothed estimator defined in (S18). More details can be found in $\mathsection$S.2 in the supplementary material. \Blue{Under Assumption \ref{ass: intensity}, according to \cite{andersen2012statistical}, $\hat{\beta}$ is $\sqrt{n}$-consistent, and 
$\hat{\lambda}_0(t)$ is a consistent estimator of $\lambda$ with rate $n^{-p/(2p + 1)}$, if the bandwidth $h_n$ satisfies
$h_n \asymp n^{-1/(2p + 1)}$ and the kernel $K$ is of order $\floor*p$, which denotes the greatest integer strictly less than $p$ \cite[p.~5]{tsybakov2009}. Under this assumption, 
the convergence rate of $\hat{\lambda}(t)$ is no slower than the uniform convergence rate given in Theorem \ref{thm-meanconsistency}. Consequently, this semiparametric estimate of $\lambda(t)$ will not affect the uniform convergence rate of the proposed mean and/or covariance function.}
The following theorem establishes the uniform convergence rate for the proposed mean function estimator.

\begin{theorem} \label{thm-meanconsistency}
Assume Assumptions \ref{asmp:idenA}--\ref{ass: intensity} hold. Then the estimated mean function $\widehat{\mu}(t) = B(t)^{\T} \widehat{\gamma}_{\mu}$, where $\widehat{\gamma}_{\mu}$ is defined in \eqref{eq-estmean},  satisfies
	\begin{equation} \label{eq-meanconsistency}
	\sup_{t \in [0, \tau] } |\widehat{\mu}(t) - \mu(t)| =  O_P\left\{q_n^{-d} + \lambda_{\mu} q_n^{m} + \left(\frac{q_n \log n}{n}\right)^{\frac{1}{2}}\right\}, 
	\end{equation}
	 provided that $\lambda_{\mu} q\lo n \hi {2m} = O(1), q_n^{\delta} = O\{({n}/{\log n})^{\delta - 2}\}$ and $\log n/n = o(q_n^{-4})$. 
\end{theorem}

\begin{remark} \label{rmk:meanest}
If $q_n \asymp (n/\log n)^{1/(1 + 2d)}$ and $\lambda_{\mu} = o\{q_n^{-(d + m)}\}$, the uniform convergence rate of $\widehat{\mu}$ is $O_P\{(n/\log n)^{-d/(1 + 2d)}\}$. Our mean function estimator achieves the optimal convergence rate $\{n/\log(n)\}^{-d/(2d+1)}$, established in \cite{stone1982} for independent and identically distributed data and in \cite{li2010uniform} for sparse functional data with the assumption that the observational times are independent of the functional data.
\end{remark}


To derive the convergence rate for the proposed covariance function estimator, we further need the following assumption.
\begin{assumption} \label{ass: covcontinuous}
	The true covariance function of $X$, $C(\cdot, \cdot)$, belongs to $C\hi d ([0, \tau] \hi 2)$. 
\end{assumption}

 Similarly to the mean function, by the result on p.149 of \cite{de1978}, Assumption \ref{ass: covcontinuous} leads the existence of  $\tilde \eta \in \R^{q_n^2}$ such that 
 $$
 \sup_{(t,s) \in [0,\tau]^2} |C(t, s) - D^{\T}(t,s)\tilde \eta| = O(q_n^{-d}).
 $$
 The following theorem establishes the convergence rate for the proposed covariance function estimator.

\begin{theorem} \label{thm-covconsistency}
	Assume Assumptions \ref{asmp:idenA}--\ref{ass: covcontinuous} hold with some $\delta > 4$ in Assumptions \ref{ass: finite4} and \ref{ass: error}. The estimated covariance function $\widehat{C}(t, s) = D(t,s)^{\T}\widehat{\eta}$, where $\widehat{\eta} = \vect(\widehat{\Xi})$ defined in \eqref{eq-estcov}, satisfies
	$$
	\sup_{(t,s) \in [0, \tau]^2 } |\widehat{C}(t,s) - C(t,s)| =O_P\left\{q_n^{-d} + \lambda_{C} q_n^{m}  + \left(\frac{q_n^2 \log n}{n}\right)^{\frac{1}{2}}\right\},
	$$
	provided that $\lambda_{C} q\lo n \hi {2m} = O(1), q_n^{\delta} = O\{({n}/{\log n})^{(\delta - 2)/2}\}$ and $\log n/n = o(q_n^{-4})$. 
\end{theorem}

\begin{remark} \label{rmk:covest}
If $q_n \asymp (n/\log n)^{1/(2d + 2)}$ and $\lambda_C = O(q_n^{-m -d})$, then the uniform convergence rate of $\hat{C}$ is $O_P\{(n/\log n)^{-d/(2d + 2)}\}$. 
In other words, the uniform convergence rate of the covariance function estimator is the same as the optimal rate established in \cite{stone1982} for independent and identically distributed data and in \cite{li2010uniform} for sparse functional data with the assumption that the observational times are independent of the functional data.
\end{remark}

\begin{corollary} \label{cor-eigen}
Under the same assumptions of Theorem \ref{thm-covconsistency}, $q_n \asymp n^{1/(4d + 2)}$ and $\lambda_C = O(q_n^{-m -d})$, for $1 \leq j \leq j_0$ satisfying $\kappa_1 > \cdots > \kappa_{j_0} > \kappa_{j_0 + 1} \geq 0$, we have
\begin{eqnarray*}
  |\widehat{\kappa}_j - \kappa_j| = O_P(n^{-\frac{1}{2}}) \quad\text{and}\quad
\left\{\int_0^{\tau}|\widehat{\varphi}_j(t) - \varphi_j(t)|^2 \de t\right\}^{\frac{1}{2}} = O_P(n^{-\frac{d}{2d + 1}}),
\end{eqnarray*}
where $\widehat{\kappa}_j$ and $\widehat{\varphi}_j$ denote the $j$th eigenvalue and eigenfunction of $\widehat{C}(t, s)$, respectively. 
\end{corollary}

This conclusion is similar to Theorem 1 of \cite{hall2006properties}, which is a refined result of Theorem 2 of \cite{yao2005}.

\section{Simulation Studies}



\subsection{Simulation design}
The simulated response process $\{X_i(t): i=1, \ldots, n\}$ is generated by $X_i(t) = \sin(t + 1/2) + \sum_{k = 1}^{50} \nu_k \zeta_{ik} \varphi_k(t) + \epsilon_i(t)$ for $t \in [0, \tau]$ with $ \tau=3$, where $\nu_k = (-1)^{k+1} (k + 1)^{-1}$, $\zeta_{ik}$'s are independently following a uniform distribution over $[-\sqrt{3}, \sqrt{3}]$ and $\varphi_k(t) = \sqrt{2/3}\cos(k \pi t) $ for $k \geq 1$ and $\epsilon_i(t)$'s are independently normally distributed across both $i$ and $t$, with mean $0$ and variance $0.01$.
We consider the following design for observation times. 
The observation times of $X_i(\cdot)$ are generated sequentially by a general counting process with the intensity function $\lambda\{t \mid \overline{X}_{i}^{\obs}(t-)\} =  \exp\{2X_i^{\obs}(t-)\}$.
This design leads to sparse observations of $X_i(t)$ with an average of $11.5$ observations on each trajectory. We vary the sample size from $n=100$ to $n=200$. 




We compare the proposed estimators with the unweighted functional principal component analysis \citep{yao2005} without adjusting for the informative observation time process. For a fair comparison, we use the penalized spline for smoothing instead of the original local linear smoother proposed by  \cite{yao2005}. For the proposed estimators, we consider both cases when the true intensity function is known or estimated to examine the impact of intensity function estimation on subsequent analysis.  
We report the mean integrated squared errors for the estimated mean function,  covariance function, and  first functional principal component, defined as $\int_0^{\tau} \{\widehat{\mu}(t) - \mu(t)\}^2 \de t$, $\int_0^{\tau} \int_0^{\tau} \{\widehat{C}(s, t) - C(s, t)\}^2 \de s\de t$, and $\int_0^{\tau} \{\widehat{\phi}_1(t) - {\phi}_1(t)\}^2 \de t$, respectively. We also report the bias and the standard error of the estimated first eigenvalue, denoted by $\widehat{\lambda}_1$.

\begin{table}[http]
\def~{\hphantom{0}}
	\centering
	\addtolength{\tabcolsep}{-2.2pt}  
	\caption{ Mean integrated squared errors ($ \times 0.01$) for the estimated mean function, covariance function, and  first functional principal component. {The actual numerical values are the ones displayed in the table multiplied by $0.01$. UW denotes the unweighted method, TW denotes the proposed method assuming that the true intensity function is known, and EW denotes the proposed method where the intensity function is estimated. }
	Standard deviations are presented in the bracket.  
	} 
	\label{tab:MISE}
	\begin{tabular}{cccccccccccccccc}
		\multirow{2}{*}{$n$}&
		\multicolumn{3}{c}{$\widehat{\mu}(t)$} & &
		\multicolumn{3}{c}{$\widehat{C}(s, t)$} & & 
		\multicolumn{3}{c}{$\widehat{\varphi}_1(t)$} & &
        \multicolumn{3}{c}{$\widehat{\kappa}_1$} \\
		
		& UW & TW & EW  & & UW & TW & EW  & & UW & TW  & EW & &  UW & TW & EW \\
		100	& 6.41 &  1.13 &  1.10  &  & 3.18 & 2.63 & 2.40 & & 14.4 & 9.8 & 9.2 & & 4.46 & 1.11 & 1.09  \\
		& (2.00) & (.67) & (.63) & & (1.55) & (1.04) & (.85) & & (8.4) & (5.3) & (4.6) & & (4.73) &(3.96) & (3.81) \\
			 & & & & & & & & & & &  \\
	200	&  6.29 &  .83 &  .82   &  & 2.94 & 1.91 & 1.75  &  & 12.6 & 7.1 & 6.6  & & 4.68 & .50 & .56  \\
	 & (1.45) & (.46) & (.42) & & (1.94) &  (.56) & (.49) & & (7.9) & (3.7) & (3.2) & & (4.02) & (3.00) & (2.91) \\

		
	\end{tabular}

\end{table}

\begin{figure}[ht]
	\centering
	{\includegraphics[width=14cm]{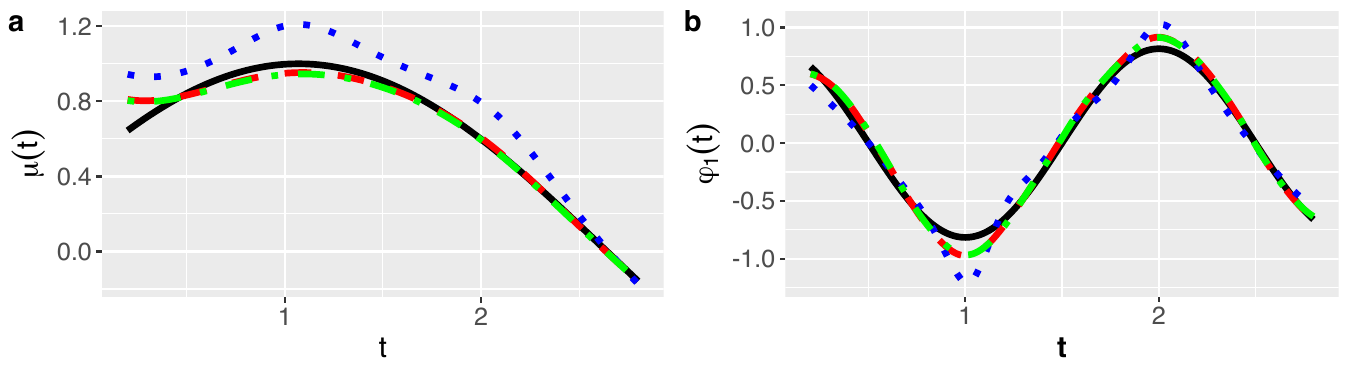}}
	\caption{
	Simulation results of the average of estimated mean function (Panel a) and the average of  estimated first functional principal component (Panel b) across 200 Monte Carlo replicates. In both panels, the black solid line denotes the true function, while the red dashed, green dash-dotted, and blue dotted lines denote the estimates from the proposed method with estimated weights, the proposed method with true weights, and the unweighted method, respectively.}
	\label{fig:FS}
\end{figure}

Table \ref{tab:MISE} summarizes mean integrated squared errors  over $200$ Monte Carlo runs. 
Figure \ref{fig:FS} plots the average of the estimated mean functions and the first functional principal components across $200$ Monte Carlo replicates.  The unweighted method shows  clear biases in estimating the mean function and the first principal component, while our proposed weighted method can reduce the biases. 
Interestingly, the proposed estimators with estimated weights improve the counterparts with true weights in terms of the mean integrated squared errors of $\widehat{C}(s,t)$ and $\widehat{\varphi}_1(t)$; see Table \ref{tab:MISE}. 
This phenomenon is similar to the inverse propensity weighting estimator of the average treatment effect, where one can achieve better efficiency by using the estimated propensity score instead of using the true score. 

In addition, we consider various designs in $\mathsection$S.3 of the supplementary material: \\
1. $\lambda(t)$ depends on both an auxiliary process $Z$ and the past history of $X$, and the true process $X$ also depends on $Z$. $Z$ can either be a null set or be a multivariate random vector or a stochastic process. \\
2. The baseline intensity function $\lambda_0(t)$ can be set to be a constant or a linear function. \\
3. The observational time is independent of the response process. \\
For all these settings, our proposed method performs similarly to the comparison shown earlier; see $\mathsection$S.3 for details.

\section{Application}

\begin{figure}[H]
	\centering
	{\includegraphics[width=14cm]{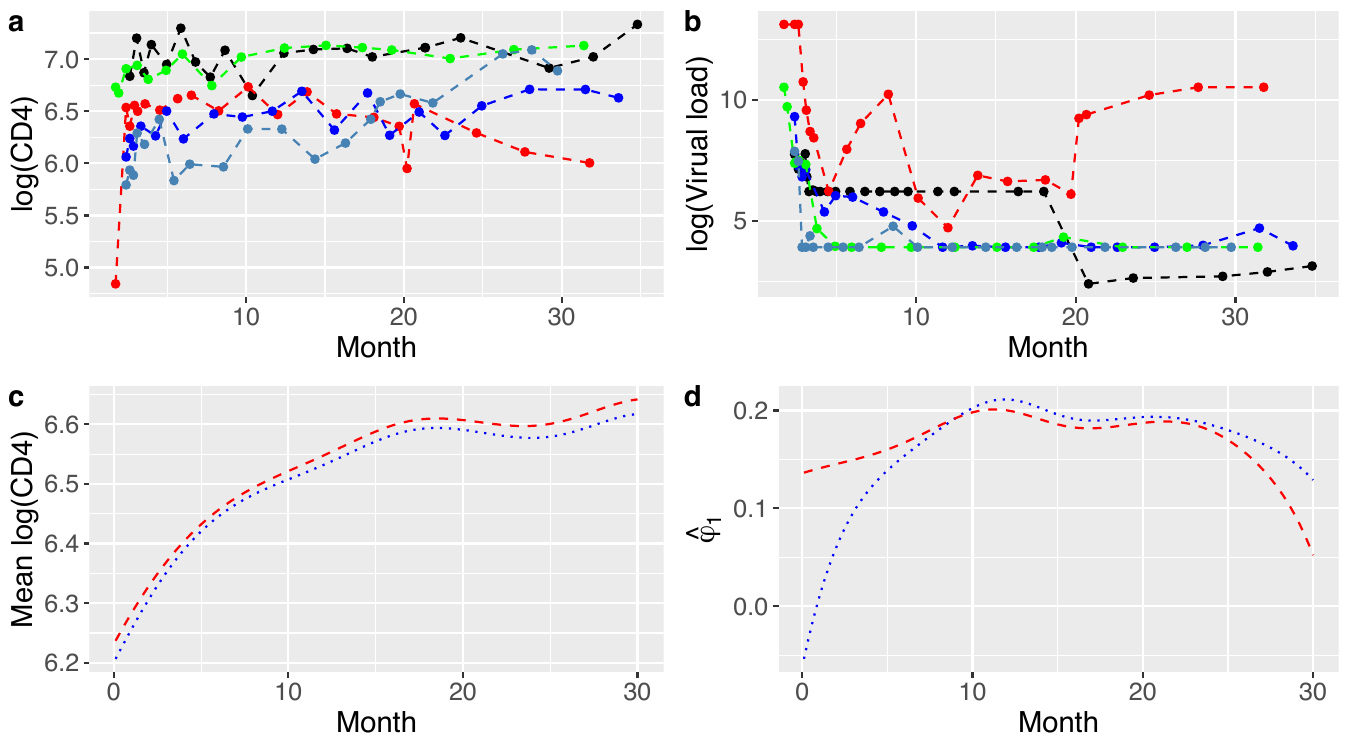}}
	\caption{(a) \& (b) Trajectories of {log} CD4 counts and log viral load observed at irregularly spaced follow-up visits from 5 randomly selected patients. (c) \& (d) Estimated mean function and the first eigenfunction  of {log} CD4 counts from the unweighted method and the proposed weighted method. Here red dashed and blue dotted lines  represent the estimates from the weighted and unweighted methods, respectively.}
\label{fig:CD4}
\end{figure}

Most existing studies focused only on the treatment effect of  highly active antiretroviral therapy on a clinical
endpoint at a fixed time point, e.g. CD4 counts at two years after
treatment initiation \citep{yang2021}. On the contrary, our goal
is to study the mean trend and variation mode of CD4 counts. 
The observational AIEDRP (Acute Infection and Early Disease Research
Program) Core 01 study was established by \cite{hecht2006}. It established
a cohort of newly infected HIV patients. The patients were protocolized
to visit the physicians for outcome assessment such as CD4 count and
viral load at weeks $2$, $4$, and $12$, and then every $12$ weeks
thereafter, through week 96. In our analysis, we include {72} patients
from the AIEDRP program who initiated HAART between $52$ and $92$
days after HIV diagnosis. These patients also had more than {$2$}
visits during the study follow-up. The outcome of interest is {log} CD4
count, lower values meaning worse immunological function. A unique
challenge arises due to substantial variability in the follow-up visit
times at which patient outcomes were assessed. Figure \ref{fig:CD4}(a)
and (b) show the trajectories of {log} CD4 counts and
log viral load at the follow-up visits from $5$ randomly selected
patients, respectively. The number and timing of visits differ from
one patient to the next, resulting in irregularly spaced observations.
Moreover, such irregular visit times can be due to obstacles that
may be related to patients' health status and thus informative about
the outcome of interest.

We apply the proposed method to estimate the mean trend and variation
mode of {log} CD4 counts over time. To address the irregularly spaced and
informative observation times, we model the intensity of visit times
by a Cox proportional intensity function adjusting for {log CD4 counts and} log viral load
at the closest past visit. The fitted result for the intensity function, presented in $\mathsection$S.4 of the supplementary material,
shows that patients with {lower CD4 counts and} higher viral load are more likely to visit.
Figure \ref{fig:CD4}(c) displays the estimated mean functions of
the {log} CD4 counts from unweighted and weighted analyses.
The unweighted estimator shows persistently lower means than the
weighted estimator over time. This is in line with the fitted result
of the intensity function which suggests that the worse outcomes are
more likely to be assessed and thus the unweighted estimator is biased
downward. Figure \ref{fig:CD4}(d) displays the estimated first eigenfunction
which depicts the dominant mode of variation of CD4 counts. The weighted
and unweighted analyses tend to agree on the variation mode after
$10$ months; however, there exist great discrepancies between them before $10$ months. 
The weighted analysis uncovers the phase transitions of CD4 counts following treatment initiation: an immediate dramatic change, followed by a plateau between $10$ months and $20$ months, and a rebound after $20$ months. Such transitions are reasonable because antiretroviral therapy promptly reduces the amount of HIV and helps recover the immune system and produce more CD4 cells, while drug resistance can be developed in extended long treatment uptake and affects CD4 counts to change.

\blue{For the sensitivity analysis of the intensity function for the observation times of CD4 counts, we fit another intensity function with $g(\overline{O}(t))$ taken as the log viral load and its square. This new intensity function leads to a similar estimate of the mean function and the first eigenfunction of log CD4 counts. More details can be found in $\mathsection$S.4 of the supplementary material.}

\section{Discussion}
To handle the informative observation time process, we describe identifying
assumptions that are tantamount to the missingness at random assumption;
that is, the unobserved outcomes are unrelated to the probabilities
of observations so long as controlling for observed information. 
Our weighting strategy can be readily extended to other functional principal component analyses, such as the Principal Analysis by Conditional Expectation proposed by \cite{yao2005}. Empirical results in $\mathsection$S.3.1 and $\mathsection$S.4 demonstrate similar performance to the proposed approach, while theoretical comparisons will be explored in future research.
More robust and efficient estimation than weighting-alone estimators can be developed by using the augmentation of the conditional mean functions (\citealp{coulombe2024multiply}), which will be another interesting future research topic.  
In practice, if a prognostic variable that is related to the observation
time process is not captured in the data, the observed information
is not sufficient to explain away the dependence between the longitudinal
outcomes and the observational time process, leading to observations
not at random or missingness not at random \citep{pullenayegum2016longitudinal,sun2021recurrent}.
Because such assumptions are untestable, sensitivity analysis methodology
is critically important for assessing the robustness of the study
conclusion against violation of assumptions; however, no such methodology
has been developed previously. In the future, we will develop a sensitivity
analysis toolkit following \citep{robins1999sensitivity,yang2017sensitivity,smith2022trials}
for functional data with irregular observation times.

\section*{Supplementary Material}

\maketitle

\setcounter{equation}{0}
\setcounter{figure}{0}
\setcounter{table}{0}
\setcounter{page}{1}
\makeatletter
\renewcommand{\theequation}{S\arabic{equation}}
\renewcommand{\thefigure}{S\arabic{figure}}
\renewcommand{\thetable}{S\arabic{table}}
\renewcommand{\bibnumfmt}[1]{[S#1]}
\renewcommand{\citenumfont}[1]{S#1}
\renewcommand{\thelemma}{S\arabic{lemma}}
\renewcommand{\thetheorem}{S\arabic{theorem}}
\renewcommand{\theproposition}{S\arabic{proposition}}

\setcounter{section}{0}
\renewcommand{\thesection}{S.\arabic{section}}

The supplementary material contains some additional details of numerical implementations, the proofs of the lemmas and theorems, and additional results of numerical studies in the main manuscript. 
\blue{More specifically, $\mathsection$\ref{Ssec:proof} contains the proof of the identification results, i.e., Equations \eqref{eq-exp}-\eqref{eq-cov}, and main theorems in the main text. 
	We present some additional details of numerical implementations in $\mathsection$\ref{Ssec:numerical}. In $\mathsection$\ref{Ssec:simulation} we perform additional simulation studies under various settings to investigate the performance of our proposed method. We present additional results of the real application in $\mathsection$\ref{Ssec:application}. 
}

\section{Proofs}
\label{Ssec:proof}
\subsection{Identification results}
In this subsection, we present detailed proofs of Equations (1)-(2) in the main text.
\begin{proof}[Proof of Equation \eqref{eq-exp}]
	\begin{align*}
	&   \quad\E[X(t)\lambda^{-1}\{t\mid\overline{O}(t)\}\de N(t)] \\
	& =\E \left(\E[X(t)\lambda^{-1}\{t\mid\overline{O}(t)\}\de N(t) \mid \F^{*}_t] \right) \quad \mbox{by the law of total expectation}\\
	&  =\E[X(t)\lambda^{-1}\{t\mid\overline{O}(t)\}\E\{\de N(t)\mid\F^{*}_t\}]~\mbox{since $X(t)$, $\overline{O}(t)$ are measurable with respect to $\F^{*}_t$ }\\
	& =\mu(t)\de t. \quad \mbox{by the definition of $\lambda\{t\mid\overline{O}(t)\}$}
	\end{align*}
\end{proof}

\begin{proof}[Proof of Equation \eqref{eq-cov}]
	\begin{eqnarray*}
		&  & \E[\{X(t)-\mu(t)\}\{X(s)-\mu(s)\}\lambda^{-1}\{s\mid\overline{O}(s)\}\de N(s)\lambda^{-1}\{t\mid\overline{O}(t)\}\de N(t)] \\
		& = & \E\left(\E[\{X(t)-\mu(t)\}\{X(s)-\mu(s)\}\lambda^{-1}\{s\mid\overline{O}(s)\}\de N(s)\lambda^{-1}\{t\mid\overline{O}(t)\}\de N(t) \mid \F^{*}_t]\right) \\ 
		& & \mbox{by the law of total expectation} \\
		& = & \E[\{X(t)-\mu(t)\}\{X(s)-\mu(s)\}\lambda^{-1}\{s\mid\overline{O}(s)\}\de N(s)\lambda^{-1}\{t\mid\overline{O}(t)\}\E\{\de N(t)\mid\F^{*}_t\}] \\
		& & \mbox{since $s < t$, $\{X(s)-\mu(s)\}\lambda^{-1}\{s\mid\overline{O}(s)\}\de N(s)$ is measurable with respect to $\F^{*}_t$ } \\
		& = & \E[\{X(t)-\mu(t)\}\{X(s)-\mu(s)\}\lambda^{-1}\{s\mid\overline{O}(s)\}\de N(s)\de t] \\
		& & \mbox{by the definition of $\lambda\{t\mid\overline{O}(t)\}$} \\
		& = & \E\left(\E[\{X(t)-\mu(t)\}\{X(s)-\mu(s)\}\lambda^{-1}\{s\mid\overline{O}(s)\}\de N(s)\de t \mid \F^{*}_s, X(t)] \right) \\
		& & \mbox{by the law of total expectation} \\
		& = & \E[\{X(t)-\mu(t)\}\{X(s)-\mu(s)\}\lambda^{-1}\{s\mid\overline{O}(s)\}\E\{\de N(s)\mid\F^{*}_s\}\de t] \\
		& & \mbox{since $\{X(t)-\mu(t)\}\{X(s)-\mu(s)\}\lambda^{-1}\{s\mid\overline{O}(s)\}$ is measurable with respect to $\sigma(X(t), \F^{*}_s)$ } \\
		& = & \E[\{X(t)-\mu(t)\}\{X(s)-\mu(s)\}\de t\de s] \\
		& & \mbox{by the definition of $\lambda\{t\mid\overline{O}(s)\}$} \\
		& = & C(t,s)\de t\de s.
	\end{eqnarray*}
\end{proof}

\subsection{Mean function estimate}

We investigate the consistency of the estimated mean function using estimated weights defined in  \eqref{eq-meanweight}.
Following \cite{lin2004analysis}, we decompose the proof into two steps: 
the first step assumes the true weights to be known, and the second step quantifies the impact of the estimated weights. 
We introduce some important notations first. For any continuous function $f$ defined on $[0, \tau]$, we define $\|f\|_{\infty} = \sup_{t \in [0, \tau]} |f(t)|$. For any two positive sequences $a_n$ and $b_n$, if there exists a constant $C_1$ such that $a_n \leq C_1 b_n$, we write $a_n \lesssim b_n$. Write $a_n \asymp b_n$ if $a_n \lesssim b_n$ and $b_n \lesssim a_n$. 

The loss function we want to minimize can be written as
\begin{equation} \label{eq-penloss}
V(\gamma) = \frac{1}{n}\sum_{i=1}^{n} \sum_{j = 1}^{m_i} \{X_{ij} - B(t_{ij})^{\T}\gamma\}^2 w_{ij}(\mu) + \lambda_{\mu}\gamma^{\T}Q_{\mu}\gamma. 
\end{equation}
We introduce some notations first to simply the following derivations. For $i = 1, \ldots, n$, let $W_i = diag(w_{i1}, \ldots, w_{im_i})$, $X_i = (X_{i1}, \ldots, X_{im_i})^{\T}$, and $B_i = \{B(t_{i1}), \ldots, B(t_{im_i})\}^{\T} \in \R^{m_i \times q_n}$. Then define $G_n = n^{-1}\sum_{i = 1}^n B_i^{\T} W_i B_i := n^{-1}B^{\T}WB$, where $W = diag(W_1, \ldots, W_n)$ and $B = (B_1^{\T}, \ldots, B_n^{\T})^{\T}  \in \R^{N \times q_n}$ with $N = \sum_{i = 1}^n m_i$. Minimizing \eqref{eq-penloss} yields 
${\gamma}^{\star} = H_n^{-1} (B^{\T} W X/n)$, where $H_n = G_n + \lambda_{\mu} Q_{\mu}$, and the corresponding $\mu^{\star}(t) = B^{\T}(t) \gamma^{\star}$. 
Next we show the uniform convergence rate of $\mu^{\star}$, which is critical to identify the order of $\|\hat{\mu} - \mu\|_{\infty}$. 

\begin{proposition} \label{thm-intermeanest-uniform}
	Assume Assumptions \ref{asmp:idenA}--\ref{ass: intensity} hold. 
	Furthermore, we assume that the knots are equally spaced in $\ca S_n$.
	Then {$\mu^*(t) = B^{\T}(t)\gamma^*$} satisfies
	\begin{equation} \label{eq: fixedest1}
	\|\mu^{*} - \mu\|_{\infty} = O(q_n^{-d}) + O(\lambda_{\mu} q_n^{m}) + O\left\{\left(\frac{q_n \log n}{n}\right)^{\frac{1}{2}}\right\}
	\end{equation}
	almost surely, 
	provided that $\lambda_{\mu} q\lo n \hi {2m} = O(1), q_n^{\delta} = O\{({n}/{\log n})^{\delta - 2}\}$ and $\log n/n = o(q_n^{-4})$. 
\end{proposition}

Before proving Proposition \ref{thm-intermeanest-uniform}, we provide the following lemmas that are useful. For simplicity, we assume $\tau = 1$.  Let $Q$ denote the uniform distribution on $[0, \tau]$. For the following lemmas, we assume that Assumptions \ref{asmp:idenA}--\ref{ass: intensity} hold. 

\begin{lemma} \label{lem-cdf}
	Let $Q_{ni}(t) = \sum_{j = 1}^{m_i} w_{ij} I(t_{ij} \leq t)$ for $i = 1, \ldots, n$ and $t \in [0, \tau]$. Then define 
	$Q_n(t) = n^{-1}\sum_{i = 1}^n Q_{ni}(t).$
	If $\log n/n = o(q_n^{-4})$, then the following result holds almost surely:
	$$
	\sup_{t \in [0, 1]} |Q_n(t) - Q(t)| = o(q_n^{-1}).
	$$
\end{lemma}

\begin{proof}[Proof of Lemma \ref{lem-cdf}]
	For $i = 1, \ldots, n$, we have 
	\begin{align*}
	\E[Q_{ni}(t) ] & = \E\left[\int_0^{\tau} w_i(t) \de N_i(t) \right] \\
	& = \int_0^{\tau} \lambda^{-1}\{t\mid\overline{O}_i(t)\} \E{\de N_i(s)} \\
	& = t.
	\end{align*}
	Define $S_{ni}(t) = Q_{ni}(t) - t$ and $S_n(t) = Q_n(t) - t$. By the definition of $Q_n$, we have $S_n(t) = n^{-1} \sum_{i = 1}^n S_{ni}(t)$. 
	As $\E[\{S_{ni}(t)\}^2]$ is bounded a common constant that is independent of $n$, $E[\{S_n(t)\}^2] \lesssim n^{-1}$ for any $t \in [0, 1]$. Then by applying the technique in the proof of Lemma A.7 in \cite{xiao2020asymptotic}, we have
	$\sup_{t \in [0, 1]} |Q_n(t) - Q(t)| = o(q_n^{-1})$ as long as $\log n/n = o(q_n^{-4})$. 
	
\end{proof}

\begin{lemma} \label{lem-G_n&G}
	Define
	
	$$
	G = \int_0^{\tau} B(s) B^{\T}(s) \de s. 
	$$
	If $\lambda_{\mu} q\lo n \hi {2m} \lesssim 1$ and $\log n/n = o(q_n^{-4})$, then the following results hold almost surely:
	\begin{itemize}
		\item[(i)] The minimal and maximal eigenvalues of $G_n$ satisfy
		$$
		\lambda_{\min}(G_n) \asymp \lambda_{\max}(G_n) \asymp q_n^{-1}.
		$$
		\item[(ii)] The minimal and maximal eigenvalues of $G$ satisfy
		$$
		\lambda_{\min}(G) \asymp \lambda_{\max}(G) \asymp q_n^{-1}.
		$$
		\item[(iii)] $\|G_n^{-1}\|_{\infty} = O(q_n)$ and $\|G^{-1}\|_{\infty} = O(q_n)$. 
		\item [(iv)] $\|G_n - G\|_{\max} = O(\|Q_n - Q\|)$.
	\end{itemize}
\end{lemma}

\begin{proof}[Proof of Lemma \ref{lem-G_n&G}]
	We rewrite $G_n$ as
	$$
	G_n = \frac{1}{n} \sum_{i = 1}^n \int_0^{\tau} \lambda^{-1}(s | \overline{O}_i(s)) B(s)B^{\T} (s) \de N_i(s).
	$$
	Part (ii) and 
	Part (iii) follow from the proof of Lemmas 6.2 and 6.3 of \cite{zhou1998local}, respectively. So the proof is omitted. 
	Regarding part (i), we rewrite $G_n$ as
	$$
	G_n = \int_0^{\tau} B(s) B^{\T}(s) \de Q_n (s).
	$$
	By Lemma \ref{lem-cdf}, we have $\|Q_n - Q\|_{\infty} = o(q_n^{-1}).$ Then we can apply integration by parts to show that part (i) holds. Therefore,
	$$
	G_n - G = \int_0^{\tau} B(s) B^{\T}(s) \de (Q_n - Q) (s). 
	$$
	As $B_k(s) \geq 0$ and $\sum_{k = }^{q_n} B_k(s) = 1$ for any $s \in [0, \tau]$, part (iv) holds by applying intergration by parts. The proof is completed. 
	
	\QEDB
\end{proof}

\begin{lemma} \label{lem-H_n&H}
	
	If $\lambda_{\mu} q\lo n \hi {2m} \lesssim 1$ and $\log n/n = o(q_n^{-4})$, then the following results hold almost surely:
	\begin{itemize}
		\item[(i)] $\|H_n^{-1}\|_{\max} = O(q_n)$, and
		\item[(ii)] $\|H_n^{-1}\|_{\infty} = O(q_n)$. 
	\end{itemize}
\end{lemma}

\begin{proof}[Proof of Lemma \ref{lem-H_n&H}]
	For part (ii), recall that
	$H_n = G_n + \lambda_{\mu}Q_{\mu}$. By Lemma A.4 of \cite{xiao2019asymptotic} and the proof of part (i) of Lemma \ref{lem-G_n&G}, we have $\|H_n^{-1}\|_{\infty} = O\{q_n (1 + \lambda_{\mu} q_n^{2m})^{3/2}\} = O(q_n)$ almost surely. For part (i), we can refer to the proof of lemma A.4 of \cite{xiao2019asymptotic}. In particular, $H_n$ is an $m^*$-banded matrix, i.e, the $(i,j)$th entry of $H_n$ is 0 if $|i - j| > m^*/2$. Furthermore, the $(k,l)$th entry of $H_n^{-1}$ satisfies
	$$
	|(H_n^{-1})_{kl}| = O\{q_n (1 + \lambda_{\mu} q_n^{2m})\} \left(\frac{\sqrt{\text{cond}(H_n)} - 1}{\sqrt{\text{cond}(H_n)} + 1}\right)^{\frac{2|k - l|}{m^*}},
	$$
	where $\text{cond}(H_n) = \lambda_{\max}(H_n)/\lambda_{\min}(H_n)$ is of order $(1 + \lambda_{\mu} q_n^{2m})$. Thus part (i) holds almost surely. The proof is completed. 
	
	\QEDB
\end{proof}

Building on Lemmas \ref{lem-cdf}--\ref{lem-H_n&H},  we prove Proposition \ref{thm-intermeanest-uniform}.

\begin{proof}[Proof of Proposition \ref{thm-intermeanest-uniform}]
	
	Denote $(t_{i1}, \ldots, t_{i,m_i})^{\T}$ by $T_i$ for $i = 1, \ldots, n$. 
	By the triangle inequality, we have
	\begin{equation} \label{eq-bias&var}
	\|\mu^{\star} - \mu\|_{\infty} \leq \|\E(\mu^{\star} \mid T_1, \ldots, T_n) - \mu\|_{\infty} + \|\mu^{\star} - \E(\mu^{\star} \mid T_1, \ldots, T_n)\|_{\infty},
	\end{equation}
	where the first term can be treated as the bias term while the second term as the variance term. 
	
	We deal with the bias term first. By Lemma 5 of \cite{stone1985}, there exists $\widetilde{\gamma} \in \R^{q_n}$ such that
	$
	\|\mu(\cdot) - B^{\T}(\cdot)\widetilde{\gamma}\|_{\infty} = O(q_n^{-d}).
	$
	Let $\mu_i = \{\mu(t_{i1}, \ldots, \mu(t_{i,m_i})\}^{\T}$ and $\underline{\mu} = (\mu_1^{\T}, \ldots, \mu_n^{\T})^{\T}$. Similarly, define $\tilde{\mu}_i = (\tilde\mu(t_{i1}, \ldots, \tilde\mu(t_{i,m_i}))^{\T}$ and $\underline{\tilde\mu} = (\tilde\mu_1^{\T}, \ldots, \tilde\mu_n^{\T})^{\T}$, where $\tilde{\mu} = B^{\T}(\cdot)\widetilde{\gamma}$.
	Since $\mu^{\star}(t) = B^{\T}(t) H_n^{-1} (B^{\T}WX)$, we have
	\begin{align}
	\begin{split} \label{eq-mean-exp}
	& \quad \E\{\mu^{\star}(t) \mid T_1, \ldots, T_n\} \\
	& = B^{\T}(t) H_n^{-1} \frac{B^{\T} W \underline\mu}{n}\\
	& = B^{\T}(t) G_n^{-1} \frac{B^{\T} W \underline\mu}{n} - 
	B^{\T}(t) H_n^{-1} (\lambda_{\mu}Q_{\mu}) H_n^{-1} \frac{B^{\T} W \underline\mu}{n} \\
	& =  B^{\T}(t) G_n^{-1} \frac{B^{\T} W (\underline\mu - \underline{\tilde\mu})}{n} + 
	B^{\T}(t) G_n^{-1} \frac{B^{\T} W \underline{\tilde\mu}}{n}
	- 
	B^{\T}(t) H_n^{-1} (\lambda_{\mu}Q_{\mu}) G_n^{-1} \frac{B^{\T} W \underline\mu}{n}.
	\end{split}
	\end{align}
	
	For this second term on the right-hand side of \eqref{eq-mean-exp}, we have 
	$$
	B^{\T}(t) G_n^{-1} \frac{B^{\T} W \underline{\tilde\mu}}{n} = B^{\T}(t) G_n^{-1} \frac{B^{\T} W B{\tilde\gamma}}{n} = B^{\T}(t) \tilde{\gamma} = \tilde{\mu}(t). 
	$$
	Therefore, 
	$$
	\E\{\mu^{\star}(t) \mid T_1, \ldots, T_n\}  - \mu(t) = (\tilde{\mu} - \mu)(t) +  B^{\T}(t) G_n^{-1} \alpha - B^{\T}(t) H_n^{-1} (\lambda_{\mu}Q_{\mu})\beta,
	$$
	where $\alpha = {B^{\T} W (\underline\mu - \underline{\tilde{\mu}})}/{n}$ and 
	$\beta = G_n^{-1}({B^{\T} W \underline\mu}/{n}) $.
	Note that $B_k(t) \geq 0$ for any $k = 1, \ldots, q_n$, and $\sum_{k = 1}^{q_n} B_k(t) = 1$ for $t \in [0, 1]$. 
	It follows that
	\begin{equation} \label{eq-bias-interm}
	\|\E\{\mu^{\star}(\cdot) \mid T_1, \ldots, T_n\} - \mu\|_{\infty} \leq \|\tilde{\mu} - \mu\|_{\infty} + \|G_n^{-1}\alpha\|_{\max} + \|H_n^{-1} (\lambda_{\mu}Q_{\mu})\beta\|_{\max}.
	\end{equation}
	Moreover, the $k$th element of $\alpha$ satisfies
	\begin{align*}
	\alpha_k & = \frac{1}{n} \sum_{i = 1}^n \sum_{j = 1}^{m_i} w_{ij} B_k(t_{ij}) \{\mu(t_{ij} - \tilde{\mu}(t_{ij})\} \\
	& = \int_0^{\tau} B_k(s) \{\mu(s) - \tilde{\mu}(s)\}\de Q_n(s) \\
	& = \int_0^{\tau} B_k(s) \{\mu(s) - \tilde{\mu}(s)\}\de Q(s) + \int_0^{\tau} B_k(s) \{\mu(s) - \tilde{\mu}(s)\}\de (Q_n - Q)(s)\\
	& := \alpha_{k1} + \alpha_{k2}.
	\end{align*}
	By Lemma 3.2 of \cite{xiao2019asymptotic}, $\max_{k} |\alpha_{k1}| = o(q_n^{-(d + 1)})$. Regarding $\alpha_{k2}$, we apply the integration by parts. By Lemma \ref{lem-cdf}, we have $\max_{k} |\alpha_{k2}| = o(q_n^{-(d + 1)})$. Therefore, $\|\alpha\|_{\max} = o(q_n^{-(d + 1)})$. It follows from Lemma \ref{lem-G_n&G} that
	\begin{equation} \label{eq-bias-term2}
	\|G_n^{-1}\alpha\|_{\max} \leq \|G_n^{-1}\|_{\infty} \|\alpha\|_{\max} = o(q_n^{-d}). 
	\end{equation}
	
	Lastly, we deal with the third term on the right-hand side of \eqref{eq-bias-interm}:  $\|H_n^{-1} (\lambda_{\mu}Q_{\mu})\beta\|_{\max}.$ 
	Write $\beta$ as
	\begin{align*}
	\beta & = G_n^{-1}\frac{B^{\T} W \underline\mu}{n} \\
	& = G_n^{-1}\frac{B^{\T} W (\underline\mu - \underline{\tilde{\mu}})}{n} + G_n^{-1}\frac{B^{\T} W \underline{\tilde{\mu}}}{n} \\
	& = \tilde{\gamma} + G_n^{-1}\frac{B^{\T} W (\underline\mu - \underline{\tilde{\mu}})}{n} \\
	& = \tilde{\gamma} + G_n^{-1} \alpha. 
	\end{align*}
	By Remark 6.1 of \cite{xiao2019asymptotic}, 
	$\|Q_{\mu}\tilde{\gamma}\|_{\max} = O(q_n^{m - 1})$. Moreover, by Proposition 4.2 of \cite{xiao2019asymptotic}, 
	$\|Q_{\mu} G_n^{-1}\alpha\|_{\max} = o(q_n^{2m -1 - d}) = o(q_n^{m - 1})$ since $m \leq d$. Thus $\|Q_{\mu}\beta\|_{\max} = O(q_n^{m - 1})$. 
	Now we can find the bound on $\|H_n^{-1} (\lambda_{\mu}Q_{\mu})\beta\|_{\max}.$ On the one hand, by Lemma \ref{lem-H_n&H}, 
	$$
	\|H_n^{-1} (\lambda_{\mu}Q_{\mu})\beta\|_{\max} \leq \lambda_{\mu} \|H_n^{-1}\|_{\infty} \|Q_{\mu}\beta\|_{\max} = O(\lambda_{\mu} q_n^m). 
	$$
	On the other hand, we have
	\begin{align*}
	\|H_n^{-1} (\lambda_{\mu}Q_{\mu})\beta\|_{\max} ^2 & \leq \beta^{\T} (\lambda_{\mu}Q_{\mu}) H_n^{-2} (\lambda_{\mu}Q_{\mu}) \beta \\
	& = O(q_n) \beta^{\T} (\lambda_{\mu}Q_{\mu}) H_n^{-1} G_n H_n^{-1} (\lambda_{\mu}Q_{\mu}) \beta.
	\end{align*}
	Since $(\lambda_{\mu}Q_{\mu}) H_n^{-1} G_n H_n^{-1} (\lambda_{\mu}Q_{\mu})  \leq \lambda_{\mu}Q_{\mu}$, 
	it follows from the above inequality that
	$$
	\|H_n^{-1} (\lambda_{\mu}Q_{\mu})\beta\|_{\max} ^2  \lesssim q_n\beta^{\T} \lambda_{\mu}Q_{\mu} \beta = O(\lambda_{\mu}q_n),
	$$
	where the last equality holds by Proposition 4.3 of \cite{xiao2019asymptotic}. 
	Combining the above results, one obtains
	\begin{equation} \label{eq-bias-term3}
	\|H_n^{-1} (\lambda_{\mu}Q_{\mu})\beta\|_{\max} = O\left[\min\{\lambda_{\mu} q_n^m, (\lambda_{\mu}q_n)^{1/2} \}\right] = O(\lambda_{\mu} q_n^{m}),
	\end{equation}
	provided that $ \lambda_{\mu} q_n^{2m} \lesssim 1$.
	
	Combining \eqref{eq-bias-interm}, \eqref{eq-bias-term2} and \eqref{eq-bias-term3}, the bias term in \eqref{eq-bias&var} satisfies
	\begin{equation} \label{eq-bias-order}
	\|\E\{\mu^{\star}(\cdot) \mid T_1, \ldots, T_n\} - \mu\|_{\infty} = O(q_n^{-d}) + O(\lambda_{\mu} q_n^{m}),
	\end{equation}
	almost surely. 
	
	Next we consider the variance term, $\|\mu^{\star} - \E(\mu^{\star} \mid T_1, \ldots, T_n)\|$,
	in \eqref{eq-bias&var}. Let $e_{ij} = X_i(t_{ij}) - \mu(t_{ij}) + \epsilon_{ij}$ for $j = 1, \ldots, m_i$ and $i = 1, \ldots, n$. Denote $(e_{i1}, \ldots, e_{i,m_i})^{\T}$ by $e_i$ and $(e_1^{\T}, \ldots, e_n^{\T})^{\T}$ by $e$. Let $H = G + \lambda_{\mu}Q_{\mu}$ with $G = \int_0^{\tau} B(s) B^{\T}(s)\de s$. 
	Define
	\begin{equation} \label{eq-rtilde}
	\tilde{r}(t) = \mu^{\star}(t) - \E(\mu^{\star}(t) \mid T_1, \ldots, T_n) = B^{\T}(t) H_n^{-1} \frac{B^{\T} W e}{n},
	\end{equation}
	and 
	$$
	r(t) = B^{\T}(t) H^{-1} \frac{B^{\T} W e}{n} = \frac{1}{n} \sum_{i = 1}^n \sum_{j = 1}^{m_i} a_{ij}(t) w_{ij} e_{ij},
	$$
	where 
	$a_{ij}(t) = B^{\T}(t) H^{-1} B(t_{ij})$. We find the bound on $\|r\|_{\infty}$ first, which helps us to find the bound on $\|\tilde{r}\|_{\infty}$ later. Let 
	$$
	L_n = \left(\frac{n q_n^{-1}}{\log n}\right)^{\frac{1}{2}}. 
	$$
	To this end, rewrite $r(t)$ as
	\begin{align} \label{eq-r(t)-decomposition}
	\begin{split}
	r(t) & = \frac{1}{n} \sum_{i = 1}^n \sum_{j = 1}^{m_i} a_{ij}(t) w_{ij} e_{ij} I(|e_{ij}| \leq L_n) + \frac{1}{n} \sum_{i = 1}^n \sum_{j = 1}^{m_i} a_{ij}(t) w_{ij} e_{ij} I(|e_{ij}| >L_n) \\
	& := r_1(t) + r_2(t). 
	\end{split}
	\end{align}
	
	For any matrix $A$, define $A_+ = (|a_{ij}|)$. 
	For the first term on the right-hand side of \eqref{eq-r(t)-decomposition}, note that under Assumptions \ref{ass: finite4} and \ref{ass: error}, applying the Holder inequality yields
	\begin{align*}
	& \frac{1}{n^2} \sum_{i = 1}^n \E \left\{\int_0^{\tau} B^{T}(t) H^{-1} B(s) \cdot \lambda^{-1}(s | \overline{O}_i(s)) e_i(s) I(|e_i(s)| \leq L_n) \de N_i(s)\right\}^2 \\
	& \lesssim \frac{1}{n} B^{\T}(t) (H^{-1})_+ {\Gamma} (H^{-1})_+ B(t),
	\end{align*}
	where $\Gamma = \sigma_{\epsilon}^2 (\gamma_{1kl})_{k, l  = 1}^{q_n} + \|C\|(\gamma_{2kl})_{k, l}^{q_n} : = \Gamma_1 + \Gamma_2$, with
	$$
	\gamma_{1kl} = \E\left\{\int_0^{\tau}\lambda^{-2}(s | \overline{O}(s)) B_k(s) B_l(s) \de N(s)\right\}
	$$
	and
	$$
	\gamma_{2kl} = \E\left\{\int_0^{\tau}\lambda^{-1}(u | \overline{O}(u)) B_k(u) \de N(u) \cdot \int_0^{\tau}\lambda^{-1}(v | \overline{O}(v)) B_k(v) \de N(v) \right\}.
	$$
	Define $M_i(t) = N_i(t) - \int_0^{t} \lambda_0(s)\exp[g\{{\overline{O}_i}(s)\}^ {\T} \beta_0] \de s$ for $i = 1, \ldots, n$. It is a martingale with respect to the filtration generated by $\sigma\{Z_i(s), X_i(s), s \leq t\}$. Furthermore, if $P_1(t)$ and $P_2(t)$ are locally bounded $\sigma(X_i(s), Z_i(s), s \leq t)$-predictable processes, we have
	\begin{equation} \label{eq-predqv}
	\E \left\{\int_0^t P_1(u) \de M_i(u) \int_0^t P_2(u)  \de M_i(u) \right\} = \E \left\{\int_0^t P_1(u)P_2(u) \lambda_0(u)  \exp[g\{{\overline{O}_i}(u)\}^{\T}\beta_0] \de u \right\}.
	\end{equation}
	By \eqref{eq-predqv}, we can show $\|\Gamma_2\|_{\max} = O(q_n^{-1})$. Additionally, $\|\Gamma_1\|_{op} = O(q_n^{-1})$. As $B_k(t) \geq 0$ and $\sum_{k = 1}^{q_n} B_k(t) = 1$ for any $t \in [0, \tau]$, by Lemma \ref{lem-H_n&H}, we have almost surely
	\begin{align} \label{eq-var-term1}
	\begin{split}
	& \frac{1}{n} B^{\T}(t) (H^{-1})_+ {\Gamma} (H^{-1})_+ B(t) \\
	& \leq \frac{1}{n}\|(H^{-1})_+ {\Gamma} (H^{-1})_+\|_{\max} \\
	& \leq \frac{1}{n} \{\sigma_{\epsilon}^2 \|H^{-1}\|_{\infty} \|H^{-1}\|_{\max} \|\Gamma\|_{op} + \|C\|_{\infty}\|H^{-1}\|_{\infty}^2 \|\Gamma_2\|_{\max}\} \\
	& = \tilde{C}_1 \frac{q_n}{n}
	\end{split}
	\end{align}
	for some positive constant $\tilde{C}_1$. Additionally, there exits some constant $\tilde{C}_2$ such that
	$$
	\sum_{j = 1}^{m_i} |a_{ij}(t) w_{ij} e_{ij} I(|e_{ij}| \leq L_n) | \leq \tilde{C}_2L_n q_n
	$$
	holds almost surely and uniformly over $i = 1, \ldots, n$. If we take $T(b)= \{n^{-b}, 2n^{-b}, \ldots, 1 - n^{-b}, 1\}$, then by the Bernstein inequality, we have for any constant $c > 0$,
	$$
	\mathbb{P} \left\{\max_{t \in T(b)} |r_1(t)| > cL_n^{-1}\right\} \leq n^b \exp\left(-\frac{c^2L_n^{-2}/2}{C^*n^{-1}q_n + c\tilde{C}_2 n^{-1}q_n/3}\right) \leq n^{b - c^*},
	$$
	where $c^*$ is constant depending on $c$, $\delta$ and $\tilde{C}_2$. By choosing a sufficiently large $c$, $c^*$ will be also be sufficiently large, and thus this series is summable. 
	By the Borel-Cantelli lemma, 
	$$
	\sup_{t \in T(b)} |r_1(t)| = O(L_n^{-1}) \quad \text{almost surely}.
	$$
	Next we consider $t \in [0, \tau] \setminus T(b)$. Note that
	$$
	\|r_1\| \leq \sup_{t \in T(b)} |r_1(t)| + \sup_{u \in T(b), v \in [0, \tau], |u - v| \leq n^{-b}} |r_1(u) - r_1(v)|. 
	$$
	For the second term on the right-hand side of the above inequality,
	\begin{align*}
	|r_1(t_1) - r_1(t_2)| & \leq \frac{1}{n} \sum_{i = 1}^n \sum_{j = 1}^{m_{i}} |a_{ij}(t_1) - a_{ij}(t_2)| w_{ij} |e_{ij}| I(|e_{ij}| \leq L_n) \\
	& \leq \max_{i, j} |a_{ij}(t_1) - a_{ij}(t_2)| \cdot \left\{\frac{1}{n} \sum_{i = 1}^n \int_0^{\tau}  \lambda^{-1}(s | \overline{O}_i(s)) (\|X_i - \mu\|_{\infty} + \|e_i\|) \de N_i(s)\right\}\\
	& = \max_{i, j} |a_{ij}(t_1) - a_{ij}(t_2)| O(1)
	\end{align*}
	almost surely, where the last equality follows from the strong law of large numbers and the Holder and Minkowski  inequalities. Moreover, 
	$$
	\sup_{t_1 \in T(b), |t_1 - t_2| \leq n^{-b}}\max_{i, j} |a_{ij}(t_1) - a_{ij}(t_2)| = O(n^{-b} q_n^2).
	$$
	Therefore, 
	$$
	\|r_1(t)\| = O(L_n^{-1})
	$$
	almost surely, 
	as long as $b$ is sufficiently large.

	Lastly, we find the bound on $r_2$ in \eqref{eq-r(t)-decomposition}. Note that almost surely and uniformly over $t \in [0, \tau]$, 
	\begin{align*}
	|r_2(t)| & \leq \frac{1}{n} \sum_{i = 1}^n \sum_{j = 1}^{m_i} |a_{ij}(t)| w_{ij} |e_{ij}| I(|e_{ij}| > L_n) \\
	& = \frac{1}{n} \sum_{i = 1}^n \int_0^{\tau} |B^{\T}(t) H^{-1} B(s)| \lambda^{-1}(s | \overline{O}_i(s)) |e_i(s)| I(|e_i(s)| > L_n) \de N_i(s) \\
	& \leq \|H^{-1}\|_{\max} \cdot \left\{ \frac{1}{n} \sum_{i = 1}^n \int_0^{\tau}  \lambda^{-1}(s | \overline{O}_i(s)) |e_i(s)| I(|e_i(s)| > L_n) \de N_i(s)\right\} \\
	& \leq O(q_n) C_0 L_n^{1 - \delta}.
	\end{align*}
	where the last inequality follows from the strong law of large numbers and the Minkowski inequality.
	Therefore, $\|r_2\|_{\infty} = O(L_n^{-1})$ since
	$
	q_n L_n^{1 - \delta } \lesssim L_n^{-1},
	$
	under the assumption 
	$
	q_n^{\delta} \lesssim ({n}/{\log n})^{\delta - 2}.
	$
	
	Combining the above results, one obtains 
	$
	\|r\|_{\infty} = O(L_n^{-1})$
	almost surely. To show that $\|\tilde{r}\|_{\infty} = O(L_n^{-1})$ almost surely, it remains to prove 
	\begin{equation} \label{eq-rdiff}
	\|r - \tilde{r}\|_{\infty} = O(L_n^{-1})
	\end{equation}
	almost surely. 
	Note that
	$$
	H_n \{(G_n + \lambda_{\mu}Q_{\mu})^{-1} - (G + \lambda_{\mu}Q_{\mu})^{-1}\} H = G - G_n. 
	$$
	It follows that
	$$
	\tilde{r}(t) - r(t) = -B^{\T}(t) H_n^{-1} (G_n - G) H^{-1} \frac{B^{T}W e}{n}. 
	$$
	Hence, by Lemmas \ref{lem-G_n&G} and \ref{lem-H_n&H}, 
	\begin{align*}
	\|\tilde{r} - r\|_{\infty} & \leq \frac{1}{n}\|H_n^{-1} (G_n - G) H^{-1} (B^{\T}W e)\|_{\max} \\
	& \leq \frac{1}{n}\|H_n^{-1}\|_{\infty} \|G_n - G\|_{\infty} \|G^{-1}\|_{\infty} \|GH^{-1} (B^{\T}W e)\|_{\max} \\
	& = o\left(\frac{q_n}{n}\right) \big\|GH^{-1} (B^{\T}W e)\big\|_{\max}
	\end{align*}
	almost surely. Note that $r(t) = n^{-1} B^{\T}(t) H^{-1} (B^{\T} W e)$. So
	$$
	\frac{1}{n}\|GH^{-1} (B^{\T}W e)\|_{\max} \leq \left\lVert\int_0^{\tau} B(s) r(s) \de s\right\rVert_{\max} = O(q_n^{-1}\|r\|),
	$$
	where the last equality holds since every $B_k(t)$ is nonzero only at a finite number of subintervals with knots as endpoints. 
	Then \eqref{eq-rdiff} holds. 
	
	Therefore $\|\tilde{r}\|_{\infty} = O(L_n^{-1})$. Combining this with \eqref{eq-rtilde}, \eqref{eq-bias-order} and \eqref{eq-bias&var}, we conclude
	$$
	\|\mu^{\star} - \mu\|_{\infty} = O(q_n^{-d}) + O(\lambda_{\mu} q_n^{m}) + O\left\{\left(\frac{q_n \log n}{n}\right)^{\frac{1}{2}}\right\}
	$$
	almost surely. The proof is completed. 
	\QEDB
\end{proof}

We are ready to prove Theorem \ref{thm-meanconsistency}. 

\begin{proof} [Proof of Theorem \ref{thm-meanconsistency}]
	From the proof of Proposition \ref{thm-intermeanest-uniform},  the minimizer of \eqref{eq-penloss}, denoted by $\gamma^{\star}$, satisfies that
	\begin{align*}
	\quad\quad U_n(\gamma^{\star}) + \lambda_{\mu}Q\gamma^{\star} 	& = -\frac{1}{n}\sum_{i = 1}^n \int_0^{\tau} \frac{\{X_i(t) - B^{\T}(t)\gamma^{\star}\}B(t)}{\lambda_0(t)\exp[g\{{\overline{O}_i}(t)\}^{\T}\beta_0]} \de N_i(t) + \lambda_{\mu} Q\gamma^{\star}\\
	& = \left(\frac{1}{n}\sum_{i = 1}^n \int_0^{\tau} \frac{ B^{\T}(t)B(t)}{\lambda_0(t)\exp[g\{{\overline{O}_i}(t)\}^{\T}\beta_0]} \de N_i(t) + \lambda_{\mu} Q \right) \gamma^{\star} \\
	& \qquad - \frac{1}{n}\sum_{i = 1}^n \int_0^{\tau} \frac{X_i(t)  B(t)}{\lambda_0(t)\exp[g\{{\overline{O}_i}(t)\}^{\T}\beta_0]} \de N_i(t) \\
	& = 0.
	\end{align*}
	It follows that
	\begin{align} \label{eq-gammsol}
	\gamma^{\star} & = \left\{\frac{1}{n}\sum_{i = 1}^n \int_0^{\tau} \frac{ B^{\T}(t)B(t)}{\lambda_0(t)\exp[g\{{\overline{O}_i}(t)\}^{\T}\beta_0]} \de N_i(t) + \lambda_{\mu} Q \right\}^{-1}  \\
	\nonumber
	& \qquad \qquad \times\left\{\frac{1}{n}\sum_{i = 1}^n \int_0^{\tau} \frac{X_i(t)  B(t)}{\lambda_0(t)\exp[g\{{\overline{O}_i}(t)\}^{\T}\beta_0]} \de N_i(t)\right\}.
	\end{align}

	Under Assumption \ref{ass: intensity}, if we apply the semi-parametric method in \cite{lin2000semiparametric} to estimate $\beta$ and the estimator of $\lambda_0(t)$ is given by \eqref{eq-lambda0-estimate}, then $\hat{\beta}$ is $\sqrt{n}$ consistent based on \cite{lin2000semiparametric}. Furthermore, by slightly adapting the proof of Theorem IV2.3 of \cite{andersen2012statistical}, we have
	$\|\hat{\lambda}_0 - \lambda_0\|_{L^2} = O_P(n^{-p/(2p + 1)}$ if $h_n \asymp n^{-1/(2p + 1)}$ and the kernel $K$ is of order $\floor*p$, which denotes the greatest integer strictly less than $p$ \cite[p.~5]{tsybakov2009}. 
	Without loss of generality, we assume that $\beta_0$ is a scalar; calculations are more involved when $\beta_0$ is a vector. 
	If we treat the right-hand side of  \eqref{eq-gammsol} as a function of $\lambda_0$ and $\beta$, denoted by $L(\lambda_0, \beta)$, then $\partial L(\lambda_0, \beta_0)/ \partial \beta $ is a vector of length $q_n$. Writing $ N_i(t) = M_i(t) + \int_0^{t} \lambda_0(s)\exp[g\{{Z_i}(s)\}^ {\T} \beta_0] \de s$,
	we can easily show that $\|\partial L(\lambda_0, \beta_0)/ \partial \beta\|_2 = O_P(1)$ and  $\|\partial L(\lambda_0, \widehat{\beta}_0)/ \partial \lambda\|_2 = O_P(1)$ under Assumptions \ref{ass: continuous} to \ref{ass: intensity}.  
	Then  it follows that
	\begin{align*}
	\widehat{\gamma} &  = L(\widehat{\lambda}_0, \widehat{\beta}) \\
	& = L(\lambda_0, \beta_0) + L(\widehat{\lambda}_0, \widehat{\beta}) - L(\lambda_0, \beta_0) \\
	& = \gamma^{\star} + L(\widehat{\lambda}_0, \widehat{\beta}) - L({\lambda}_0, \widehat{\beta})+  L({\lambda}_0, \widehat{\beta}) - L(\lambda_0, \beta_0) \\
	& = \gamma^{\star}  + \frac{\partial L(\lambda_0, \widehat{\beta})}{\partial \lambda} (\widehat{\lambda} -\lambda_0) + \frac{\partial L(\lambda_0, \beta_0)}{\partial \beta}(\widehat{\beta} - \beta) + o_P\left(n^{-\frac{p}{2p + 1}}\right).
	\end{align*}
	Therefore, the resulting estimator $\widehat{\gamma}$ satisfies that 
	$$
	\|\widehat{\gamma} - \gamma^{\star}\|_2 = O_P\left(n^{-\frac{p}{2p + 1}}\right).
	$$
	Since $\widehat{\mu}(t) = B^{\T}(t)\widehat{\gamma}$, by Theorem \ref{thm-intermeanest-uniform}, \eqref{eq-meanconsistency} holds. The proof is completed. 
\end{proof}

\subsection{Covariance function estimate}

{For simplicity of presentation, we assume that the true mean function $\mu(t)$ is known when estimating the covariance function. The results can be easily extended to the case when the mean function is estimated by the proposed method at a cost of additional technical complexity. }

With a slight abuse of notation, define $e_{ij} = X_{ij} - \mu(t_{ij})$ and $\tilde{\sigma}_{ij_1j_2} = e_{ij_1}e_{ij_2}$ for pseduo covariance between $X(t_{ij_1})$ and $X(t_{ij_2})$. Let
$\Sigma_{i} = \{C(t_{ij_1, t_{ij_2}})\}_{j_1, j_2 = 1}^{m_i} \in \R^{m_i \times m_i}$ and $\sigma_i = \vect^*(\Sigma_i)$, where $\vect^*$ denotes a matrix operator similar to $\vect$ except that it removes the diagonal elements of the matrix. Likewise, let
$\hat{\sigma}_i = (\tilde{\sigma}_{ij_1j_2}) \in \R^{m_i \times m_i} $ and $\hat{\sigma}_i = \vect^*(\hat{\Sigma}_i)$. Similar to the mean estimation, define $B_i = \{B(t_{i1}, \ldots, B(t_{im_i})\}^{\T} \in \R^{m_i \times q_n}$ and let $A_i$ be the submatrix of $B_i \otimes B_i$ with rows having the same $t_{ij}$'s removed. Thus, $A_i \in \R^{\{m_i \times (m_i - 1)\} \times q_n^2}$. Lastly, we define the weight matrix $W_i \in \R^{\{m_i \times (m_i - 1)\} \times \{m_i \times (m_i - 1)\}}$, which is a diagonal matrix with entries $w_{ij_1j_2} = w_{ij_1}w_{ij_2}$
for $j_1 \neq j_2$.

With notations defined in the last paragraph, we rewrite the optimization problem \eqref{eq-estcov} as
\begin{equation} \label{eq-covobjective}
V(\eta) = \frac{1}{n}\sum_{i = 1}^n (\hat{\sigma}_i - A_i \eta)^{\T} W_i (\hat{\sigma}_i - A_i \eta)   + \lambda_C \eta^{\T} {Q}_{C} \eta,
\end{equation}
where the estimated weights are replaced by the true ones. 
Let ${\eta^*}$ denote the minimizer of $\eqref{eq-covobjective}$:
\begin{equation} \label{eq-pcovest}
\eta^* = \argmin_{\eta \in \R\hi{q\lo n ^ 2}} V(\eta). 
\end{equation}
Let $\hat{\sigma} = (\hat{\sigma}_1^{\T}, \ldots, \hat{\sigma}_n^{\T})^{\T}$, $A = (A_1^{\T}, \ldots, A_n^{T})^{\T}$ and $W = diag(W_1, \ldots, W_n)$. Further define $G_{C, n} = \sum_{i = 1}^n A_i^T W_i A_i = AWA$ and $H_{C, n} = G_{C,n} + \lambda_C Q_C$. Then we have
$$
\eta^* = H_{C, n}^{-1} \frac{A^{\T}W \hat{\sigma}}{n}.
$$
The following proposition establishes the consistency and convergence rate of {$C^*(t, s)$}.
\begin{proposition} \label{thm-intercovest-uniform}
	Assume Assumptions \ref{asmp:idenA}--\ref{ass: covcontinuous} hold with some $\delta > 4$ in Assumptions \ref{ass: finite4} and \ref{ass: error}. 
	Furthermore, we assume that the knots are equally spaced in $\ca S_n$.
	Then $C^{*}(t, s) = D(t,s)^{\T}\eta^*$, where $\eta^*$ is the solution to \eqref{eq-pcovest}, satisfies
	$$
	\sup_{(t,s) \in [0, \tau]^2 } |C^{*}(t, s) - C(t, s)| = O(q_n^{-d}) + O(\lambda_{C} q_n^{m}) + O\left\{\left(\frac{q_n^2 \log n}{n}\right)^{\frac{1}{2}}\right\}
	$$
	almost surely, 
	provided that $\lambda_{C} q\lo n \hi {2m} = O(1), q_n^{\delta} = O\{({n}/{\log n})^{(\delta - 2)/2}\}$ and $\log n/n = o(q_n^{-4})$. 
\end{proposition}

To show this proposition, we lay out several important lemmas, as has been done for mean function estimation. For the following theorems, we assume that Assumptions \ref{asmp:idenA}--\ref{ass: covcontinuous} hold. Define $M_i(t) = N_i(t) - \int_0^{t} \lambda_0(s)\exp[g\{{\overline{O}_i}(s)\}^ {\T} \beta_0] \de s$ for $i = 1, \ldots, n$. It is a martingale with respect to the filtration generated by $\sigma\{Z_i(s), X_i(s), s \leq t\}$.

\begin{lemma} \label{lem-G_nc&G}
	Define
	$$
	G_C = \int_0^{\tau} \int_0^{\tau} D(t, s)^{\otimes 2} \de t \de s. 
	$$
	If $\lambda_{C} q\lo n \hi {2m} \lesssim 1$, $q_n = o(n^{1/2})$ and $\log n/n = o(q_n^{-4})$, then the following results hold almost surely.
	\begin{itemize}
		\item[(i)] The minimal and maximal eigenvalues of $G_n$ satisfy
		$$
		\lambda_{\min}(G_{C,n}) \asymp \lambda_{\max}(G_{C,n}) \asymp q_n^{-2}. 
		$$
		\item[(ii)] The minimal and maximal eigenvalues of $G_C$ satisfy
		$$
		\lambda_{\min}(G_C) \asymp \lambda_{\max}(G_C) \asymp q_n^{-2}.
		$$
		\item[(iii)] $\|G_{C,n}^{-1}\|_{\infty} = O(q_n^2)$ and $\|G_C^{-1}\|_{\infty} = O(q_n^2)$. 
		\item [(iv)] $\|G_{C,n} - G_C\|_{\max} = o(q_n^{-2})$.
	\end{itemize}
\end{lemma}

\begin{proof} [Proof of Lemma \ref{lem-G_nc&G}]
	
	
	We first conduct eigen-analysis of $G_C$. By Theorem 5.4.2 of \cite{devore1993},
	there exists some constant $C_0$ such that 
	\begin{equation} \label{eq-eigvD2cov}
	C_0^{-1}q_n^{-2} \leq \lambda_{\min}(G_C) \leq \lambda_{\max}(G_C) \leq C_0q_n^{-2}.
	\end{equation}
	
	For part (i), note that $G_{C,n}$ can be written as
	$$
	G_{C,n} = \frac{1}{n}\sum_{i = 1}^n \int_0^{\tau} \int_0^{\tau} \frac{D(t, s)^{\otimes 2}}{\lambda(t)\lambda(s)} \de M_i(t) \de M_i(s) +  \int_0^{\tau} \int_0^{\tau} D(t, s)^{\otimes 2} \de t \de s
	$$
	By the strong law of large numbers, the first term converges to 0 almost surely. That is part (i) holds almost surely. To prove part (iii), we only need to show that $\|G_C^{-1}\|_{\infty} = O(q_n^{-2})$. Since $G_C$ is band matrix, we apply Theorem 2.2 of \cite{demko1977inverses} to prove this statement. Let $\lambda_{\max}$ denote the largest eigenvalue of $G_C$. Note $G_C$ is positive semidefinite. Therefore, $\|\lambda_{\max}^{-1} G_C\|_2 \leq 1$. On the other hand, by part (ii), $\|\lambda_{\max}^{-1} G_C\|_2\| = \lambda_{\max}/\lambda_{\min}(G_C) \asymp 1$. Then by Theorem 2.2 of \cite{demko1977inverses}, there must exist some positive constant $\iota \in (0, 1)$ such that
	$\lambda_{\max} |(G_C^{-1})_{ij}| = O(\iota^{|i - j|})$. Therefore,  $|(G_C^{-1})_{ij}| \lesssim q_n^2\iota^{|i - j|}$. From here we can easily derive (iii). Proof of part (iv) is similar to that of Lemma \ref{lem-G_n&G}, and thus is omitted. This completes the proof. 
	\QEDB
\end{proof}

\begin{lemma} \label{lem-H_nc&H}
	If $\lambda_{C} q\lo n \hi {2m} \lesssim 1$ and $\log n/n = o(q_n^{-4})$, then the following results hold almost surely:
	\begin{itemize}
		\item[(i)] $\|H_{C,n}^{-1}\|_{\max} = O(q_n^2)$, and
		\item[(ii)] $\|H_{C,n}^{-1}\|_{\infty} = O(q_n^2)$.
	\end{itemize}
\end{lemma}

\begin{proof}[Proof of Lemma \ref{lem-H_nc&H}]
	It suffices to prove part (ii) as $\|H_{C,n}^{-1}\|_{\max} \leq \|H_{C,n}^{-1}\|_{\infty}$. 
	For part (ii),  we can apply the same method in proving Lemma \ref{lem-G_nc&G} to prove it. In particular, by Lemma A.11 of \cite{xiao2019a}, $\lambda_{\min}(H_{n, C}) \asymp \lambda_{\max}(H_{n, C}) \asymp q_n^{-2}$. Then note that $H_{n, C}$ is also a banded matrix. Thus we apply Theorem 2.2 of \cite{demko1977inverses} to verify this conclusion. 
	
	\QEDB
\end{proof}

By Lemmas \ref{lem-G_nc&G} and \ref{lem-H_nc&H}, Proposition \ref{thm-intercovest-uniform} and Theorem \ref{thm-covconsistency} can be shown based on similar arguments in Proposition \ref{thm-intermeanest-uniform} and Theorem \ref{thm-meanconsistency}.

\begin{proof}[Proof of Corollary \ref{cor-eigen}]
	We mainly follow the proof of Theorem 1 in \cite{hall2006properties} to verify this assertion. 
	
	For any function $\beta$ defined on $[0, \tau]^2$, write $\|\beta\| = (\int_0^{\tau}\int_0^{\tau} \beta(s, t)^2 \de s \de t)^{1/2}$ and for any $j \in \mathbb{N}$,
	$$
	\|\beta\|_{(j)}^2 = \int_0^{\tau} \left\{\int_0^{\tau} \beta(u, v) \varphi_j(v) \de v\right\}^2 \de u. 
	$$
	Let $\Delta = \hat{C} - C$ and $\Delta_1$ be a quantity defined as equation (4.17) in \cite{hall2006properties}.
	Then equation (4.23) of \cite{hall2006properties}, we have
	$$
	\|\hat{\varphi}_j - \varphi_j\|^2 = \kappa_j^{-2} \E^{\prime}\|\Delta_1\|_{(j)}^2 + h^4 a_j + o_P\{(nh)^{-1} + h^4\},
	$$
	where $h = q_n^{-2}$, $a_j$ is a constant that does not depend on $n$, and $\E^{\prime}$ denotes the conditional expectation given observation times. After tedious calculations as in step (V) of \cite{hall2006properties}, we obtain
	$$
	\E^{\prime}\|\Delta_1\|_{(j)}^2 = O\{(nh)^{-1}\}. 
	$$
	Therefore, if $q_n \asymp n^{1/(4d + 2)}$, we obtain the desired convergence rate for $\hat{\varphi}_j$. 
	
	Regarding the convergence rate of the estimated eigenvalue, $\hat{\kappa}_j$, by equation (4.24) of \cite{hall2006properties}, we have
	$$
	\hat{\kappa}_j - \kappa_j = \int_0^{\tau}\int_0^{\tau} (\Delta_1 - 2\Delta_4) \varphi_j(u) \varphi_j(v) \de u \de v + o_P(n^{-1/2}). 
	$$
	Moreover, by the definition of $\Delta_1$ and $\Delta_4$ given in equation (4.17) of \cite{hall2006properties}, we can easily see that they are asymptotically equal to average of i.i.d terms with finite variances. That is why we can achieve a parametric convergence rate for $\hat{\kappa}_j$.

	\QEDB
\end{proof}

\section{Numerical Implementations}
\label{Ssec:numerical}

\subsection{Estimation of the baseline intensity function}
In practice, to avoid misspecification of a parametric form for the baseline function, 
we follow the approach proposed by \cite{lin2004analysis} to estimate $\lambda_0(t)$ nonparametrically. In particular, we can use the partial likelihood approach to estimate $\beta$, denoted by $\widehat{\beta}$, and estimate the cumulative baseline intensity function $\Lambda_0(t) = \int_0^t \lambda_0(s) \de s$ using the Breslow's estimator:
$$
\widehat{\Lambda}_0(t) = \int_0^t \frac{\sum_{i = 1}^n \de N_i(s)}{\sum_{i = 1}^n \exp[g\{\overline{O}_i(s)\}^{\T}\widehat{\beta]}}.
$$
We can further consider a kernel-smoothed estimator of $\lambda_0(t)$, which is given by
\begin{equation} \label{eq-lambda0-estimate}
\widehat{\lambda}_0(t) = \frac{1}{h_n} \int_0^{\tau} K\left(\frac{t - s}{h_n} \right)\de \widehat{\Lambda}_0(s).
\end{equation}
Here $h_n$ denotes the bandwidth depending on sample size and $K(\cdot)$ is a kernel.

\blue{In addition to the Cox proportional model for the intensity function, 
	other models for $\lambda(t \mid \overline{O}_i(t))$ include the additive Aalen model \citep{aalen1980}, the additive and multiplicative model \citep{scheike2002}, the accelerated failure
	time model \citep{lin1998} and a transformed Cox model \citep{zeng2006}.}

\subsection{Choosing the number of knots} \label{Ssec:knots}
When we apply the penalized splines to estimate both the mean function and the covariance function, we need to choose the number of knots, which is denoted by $q_n$ in the main text. Here we take mean estimation as an example to illustrate our scheme for selecting $q_n$.

By the theoretical analysis, we know $q_n = o(n^{1/2})$ but $q_n$ diverges to infinity as $n \rightarrow \infty$. As argued in \cite{ruppert2002}, compared with the smoothing parameter $\lambda_{\mu}$ (or $\lambda_C$), choosing the number of knots (or basis functions) 
is less important, as long as it is sufficiently large to capture the feature of the mean function (or the covariance function). He justified this conclusion through empirical studies. 
In most cases, increasing this number beyond the necessary one has little effect on the fit. However, there may be exceptions.
We adopt the strategy proposed by \cite{ruppert2002}. More specifically, 
we propose a sequence of candidate values for $q_n$ (e.g., $5,10,20,$ so on) while ensuring that the largest value does not exceed the number of aggregated observations used to estimate the mean function. For each $q_n$ in the sequence, we use the generalized cross-validation criterion, denoted by GCV $V_\mu(\lambda_\mu)$ in the main text, to select the best $\lambda_\mu$. We prespecify a value for $\alpha$ (e.g., $0.98$) and compare the GCV at $q_n=10$ with $\alpha$ times the GCV at $q_n=5$. If the GCV at $q_n = 10$ is greater than $\alpha$ times the GCV at $q_n = 5$, then we choose $q_n = 5$. Otherwise, we proceed to the next value of $q_n$ and compare the GCV at $q_n = 10$ and $20$ using the same criteria.

\subsection{Choice of the covariate process} \label{Ssec:Z}
In our proposed model, the covariate process $Z_i(t)$ plays a critical role in describing the dependence between the observation times $\{t_{ij}: j = 1, \ldots, N_i\}$ and the response process $\{X_{ij}: j = 1, \ldots, N_i\}$. 
The covariate process $Z_i(t)$ includes all auxiliary information that is both predictive of the intensity function of the observational times and is predictive of the outcome process. It is allowed to be observed at different time points from $X_i(t)$. In practice,
$Z_i(t)$ can be time-invariant, multi-dimensional, or time-varying. In $\mathsection$ \ref{Ssec:simulation}, we consider simulation settings where $Z_i(t)$ is a time-invariant vector and/or a stochastic process. The results suggest that our proposed method compares favorably with the conventional method whether $Z_i(t)$ is a time-invariant or time-varying process. 

For practitioners, 
if they believe there exist any covariates that are both predictive of the observational times of $X_i$ and related to the value of $X_i(t)$, these covariates should be included in $Z_i(t)$. 

\section{Additional results of simulation studies}
\label{Ssec:simulation}
In this section, we consider various settings for the baseline intensity function $\lambda_0(t)$ and the auxiliary covariate process $Z(t)$ to demonstrate the finite-sample performance. The duration time for the observations is fixed at $[0, 3]$.

\subsection{Comparing the weighted methods in Setting 1}
\label{Ssec:simFPCA}

We apply the proposed weights to the method of Principal Analysis by Conditional Expectation. 
Table \ref{tab:PACE_setting1} summarizes the results of estimating the mean function, the covariance function, and the first functional principal component. By comparing Table \ref{tab:PACE_setting1} and Table \ref{tab:MISE} in the main text, we find there exists little difference between these two weighted methods in estimating the mean function, the covariance function, and the first functional principal component. 

\begin{table}[http]
	\def~{\hphantom{0}}
	\centering
	\addtolength{\tabcolsep}{12pt}  
	\caption{ Mean integrated squared errors ($ \times 0.01$) for the estimated mean function, covariance function, and  first functional principal component based on the method of Principal Analysis by Conditional Expectation in Setting 1. {The actual numerical values are the ones displayed in the table multiply by $0.01$.}
		Standard deviations are presented in the bracket.  
	} 
	\label{tab:PACE_setting1}
	\begin{tabular}{cccc}
		$n$& $\widehat{\mu}(t)$ & $\widehat{C}(s, t)$ & $\widehat{\varphi}_1(t)$ \\
		100	& 1.06 &  2.55 &  9.03   \\
		& (.68) & (.91) & (4.69)  \\
		& & & \\
		200	&  .81 &  2.33 &  7.13    \\
		& (.48) & (.85) & (3.74)  \\

		
	\end{tabular}

\end{table}


\subsection{Setting 2: $\lambda_0(t)$ is linear, and $Z$ vanishes}
\begin{figure}[H]
	\centering
	{\includegraphics[width=14cm]{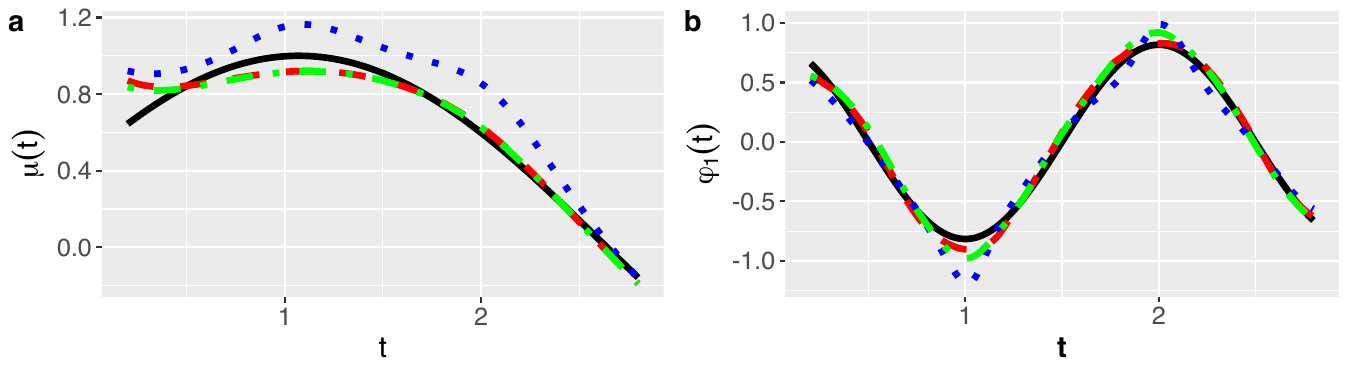}}
	\caption{
		Simulation results of the average of estimated mean function (Panel a) and the average of  estimated first functional principal component (Panel b) across 200 Monte Carlo replicates in Setting 2. In both panels, the black solid line denotes the true function, while the red dashed, green dash-dotted, and blue dotted lines denote the estimates from the proposed method with estimated weights, the proposed method with true weights, and the unweighted method, respectively.}
	\label{fig:FS_S2}
\end{figure}

The response process is simulated in the same manner as in Setting 1, while the intensity function is given by $\lambda\{t \mid \overline{O}_{i}(t)\} =  (t + 1)/2 \cdot \exp\{2X_i^{\obs}(t-)\}$. Consequently, there are approximately 11 observations per curve. Once again, the covariate process $Z_i(t)$ vanishes in this setting. Table \ref{tab:MISE_setting2} summarizes the mean integrated squared errors for the estimated mean function, covariance function, and the first functional principal component; Figure \ref{fig:FS_S2} depicts the average of the estimated mean functions and first functional principal component through 200 Monte Carlo replicates.

\begin{table}[http]
	\def~{\hphantom{0}}
	\centering
	\addtolength{\tabcolsep}{-0.1pt}  
	\caption{ Mean integrated squared errors ($ \times 0.01$) for the estimated mean function, covariance function, and  first functional principal component in Setting 2. {The actual numerical values are the ones displayed in the table multiply by $0.01$.}
		Standard deviations are presented in the bracket. UW denotes the unweighted method, TW denotes the proposed method assuming that the true intensity function is known, and EW denotes the proposed method where the intensity function is estimated. 
	} 
	\label{tab:MISE_setting2}
	\begin{tabular}{cccccccccccc}
		\multirow{2}{*}{$n$}&
		\multicolumn{3}{c}{$\widehat{\mu}(t)$} & &
		\multicolumn{3}{c}{$\widehat{C}(s, t)$} & & 
		\multicolumn{3}{c}{$\widehat{\varphi}_1(t)$}\\
		
		& UW & TW & EW  & & UW & TW & EW  & & UW & TW  & EW  \\
		100	& 7.31 &  1.42 &  1.59  &  & 3.76 & 2.51 & 3.25 & & 17.0 & 10.4 & 14.5  \\
		& (2.45) & (.95) & (.95) & & (2.65) & (.68) & (1.76) & & (9.8) & (5.6) & (11.9) \\
		& & & & & & & & & & & \\
		200	&  7.28 &  1.13 &  1.36   &  & 3.51 & 1.89 & 2.68  &  & 14.6 & 7.1 & 10.6   \\
		& (1.79) & (.59) & (.75) & & (2.98) &  (.43) & (1.94) & & (9.5) & (3.7) & (9.0)  \\
	\end{tabular}

\end{table}

\subsection{Setting 3: $\lambda_0(t)$ is a constant, and $Z$ is a random process}
In this setting, the functional response is generated as $X_i(t) = \sin(t + 1/2) - Z_i(t) + \sum_{k = 1}^{50} \nu_k \zeta_{ik} \varphi_k(t) + \epsilon_i(t)$, where $Z_i(t)$ is a Gaussian process with mean $-1.5 + 0.066t$ and covariance $0.04\delta_{st}$.  
The intensity function for the observation times is given by $\lambda\{t \mid \overline{O}_{i}(t)\} =  1/10 \cdot \exp\{0.75X_i^{\obs}(t-) - 1.5Z_i^{\obs}(t-)\}$. Consequently, there are 
approximately 11 observations per curve. 

As before, we summarize the mean integrated squared errors for the estimated mean function, covariance function, and the first functional principal component in Table \ref{tab:MISE_setting3}. Additionally, Table \ref{tab:MISE_setting3} includes the results from 
the proposed weighted method with a misspecified intensity function that ignores $Z_i(t)$ in the Cox model. We find that when a misspecified intensity function would lead to a biased estimate of the mean function when implementing the proposed method. In contrast, such misspecification has little impact on the estimated 
first corresponding eigenfunction, as shown in Table \ref{tab:MISE_setting3}.

\begin{table}[http]
	\def~{\hphantom{0}}
	\centering
	\addtolength{\tabcolsep}{-1.5pt}  
	\caption{ Mean integrated squared errors ($ \times 0.01$) for the estimated mean function, covariance function, and  first functional principal component in Setting 3. {The actual numerical values are the ones displayed in the table multiplied by $0.01$.}
		Standard deviations are presented in the bracket. UW denotes the unweighted method, TW denotes the proposed method assuming that the true intensity function is known, EW denotes the proposed method where the intensity function is estimated, and MW denotes the proposed weighted method with a misspecified intensity function that ignores $Z_i(t)$ in the Cox model.
	} 
	\label{tab:MISE_setting3}
	\begin{tabular}{ccccccccccccccc}
		\multirow{2}{*}{$n$}&
		\multicolumn{4}{c}{$\widehat{\mu}(t)$} & &
		\multicolumn{4}{c}{$\widehat{C}(s, t)$} & & 
		\multicolumn{4}{c}{$\widehat{\varphi}_1(t)$} \\
		
		& UW & TW & EW& MW   & & UW & TW & EW & MW  & & UW & TW  & EW & MW \\
		100	& 1.18 &  .41 &  .38 & .68  &  & 1.34 & 2.30 & 2.17 & 1.53 & & 6.99 & 5.45 & 5.28 & 6.38   \\
		& (.45) & (.24) & (.24) & (.34) & & (.26) & (.50) & (.48) & (.29) & & (4.33) & (2.38) & (2.30) & (3.36) \\
		& & & & & & & & & & & & & &  \\
		200	&  1.02 &  .29 &  .27 & .58   &  & 1.02 & 1.92 & 1.80 & 1.18  &  & 4.77 & 4.16 & 4.21 & 4.29   \\
		& (.30) & (.15) & (.15) & (.25) & & (.15) &  (.33) & (.30)& (.15) & & (2.83) & (1.60) & (1.57) & (1.90) \\

		
	\end{tabular}

\end{table}

\begin{figure}[H]
	\centering
	{\includegraphics[width=14cm]{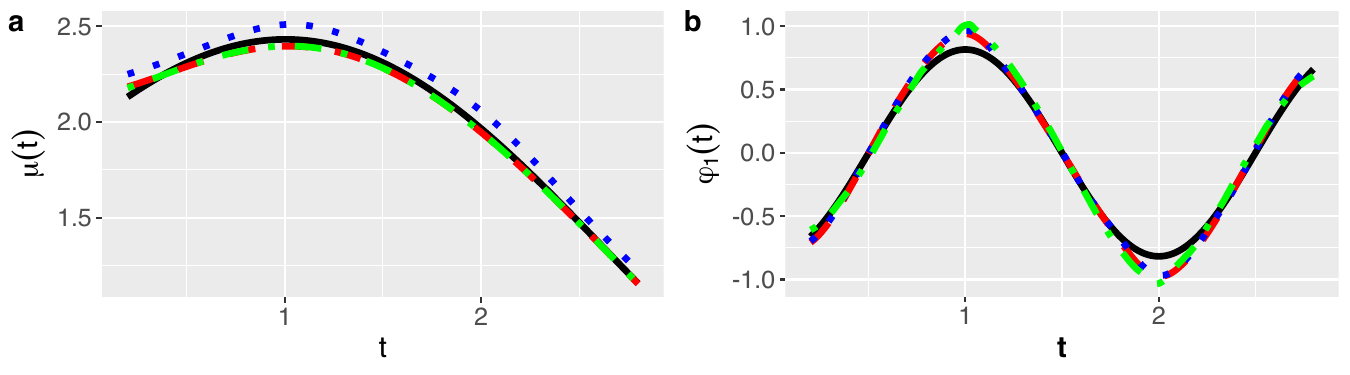}}
	\caption{
		Simulation results of the average of estimated mean function (Panel a) and the average of  estimated first functional principal component (Panel b) across 200 Monte Carlo replicates in Setting 3. In both panels, the black solid line denotes the true function, while the red dashed, green dash-dotted, and blue dotted lines denote the estimates from the proposed method with estimated weights, the proposed method with true weights, and the unweighted method, respectively.}
	\label{fig:FS_S3}
\end{figure}

\subsection{Setting 4: $\lambda_0(t)$ is linear, and $Z$ is a random process}
In this setting, the function response is generated as $X_i(t) = \sin(t + 1/2) - Z_i(t) + \sum_{k = 1}^{50} \nu_k \zeta_{ik} \varphi_k(t) + \epsilon_i(t)$, where $Z_i(t)$ is a Gaussian process with mean $-1.5 + 0.066t$ and covariance $0.04\delta_{st}$.  
The intensity function is given by $\lambda\{t \mid \overline{O}_{i}(t)\} =  (t + 1)/50 \cdot \exp\{0.75X_i^{\obs}(t-) - 3Z_i^{\obs}(t-)\}$. Consequently, there are 
approximately 32 observations per curve. This can be treated as a setting with dense observations. To save computational cost, we consider $n = 50$ and 100 curves, respectively.  

Table \ref{tab:MISE_setting4}
the mean integrated squared errors for the estimated mean function, covariance function, and the first functional principal component  for the methods introduced in the main text, as well as the result from the proposed weighted method with a misspecified intensity function that ignores $Z_i(t)$ in the Cox model. We find that the estimated mean function from the proposed method with a correctly specified intensity function is considerably more accurate than those from ignoring the weight or misspecifying the intensity function.

\begin{table}[http]
	\def~{\hphantom{0}}
	\centering
	\addtolength{\tabcolsep}{-1.5pt}  
	\caption{ Mean integrated squared errors ($ \times 0.01$) for the estimated mean function, covariance function, and  first functional principal component in Setting 4. {The actual numerical values are the ones displayed in the table multiplied by $0.01$.}
		Standard deviations are presented in the bracket.  UW denotes the unweighted method, TW denotes the proposed method assuming that the true intensity function is known, EW denotes the proposed method where the intensity function is estimated, and MW denotes the proposed weighted method with a misspecified intensity function that ignores $Z_i(t)$ in the Cox model.
	} 
	\label{tab:MISE_setting4}
	\begin{tabular}{ccccccccccccccc}
		\multirow{2}{*}{$n$}&
		\multicolumn{4}{c}{$\widehat{\mu}(t)$} & &
		\multicolumn{4}{c}{$\widehat{C}(s, t)$} & & 
		\multicolumn{4}{c}{$\widehat{\varphi}_1(t)$} \\
		
		& UW & TW & EW& MW   & & UW & TW & EW & MW  & & UW & TW  & EW & MW \\
		50	& 1.75 &  .65 &  .69 & 2.00  &  & 1.68 & 3.18 & 2.00 & 1.71 & & 10.19 & 8.53 & 9.68 & 9.29   \\
		& (.45) & (.24) & (.24) & (.81) & & (.26) & (.50) & (.48) & (.44) & & (7.33) & (5.26) & (5.81) & (7.71) \\
		& & & & & & & & & & & & & &  \\
		100	&  1.56 &  .40 &  .45 & 1.80   &  & 1.17 & 2.45 & 1.38 & 1.24  &  & 6.40 & 5.79 & 6.01 & 6.17   \\
		& (.47) & (.29) & (.33) & (.63) & & (.25) &  (.56) & (.65)& (.26) & & (4.00) & (2.47) & (3.24) & (3.58) \\

		
	\end{tabular}

\end{table}

\begin{figure}[H]
	\centering
	{\includegraphics[width=14cm]{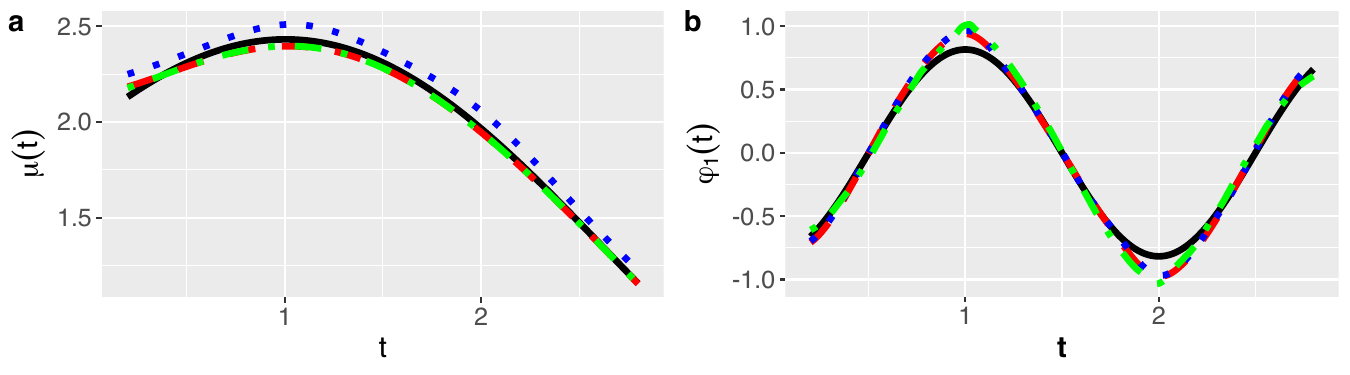}}
	\caption{
		Simulation results of the average of estimated mean function (Panel a) and the average of  estimated first functional principal component (Panel b) across 200 Monte Carlo replicates in Setting 4. In both panels, the black solid line denotes the true function, while the red dashed, green dash-dotted, and blue dotted lines denote the estimates from the proposed method with estimated weights, the proposed method with true weights, and the unweighted method, respectively.}
	\label{fig:FS_S4}
\end{figure}

\subsection{Setting 5: $\lambda_0(t)$ is linear, and $Z$ is a random vector}
The functional response is generated from
$X_i(t) = \sin(t + 1/2) - 0.25Z_1 + 0.15Z_2 + \sum_{k = 1}^{50} \nu_k \zeta_{ik} \varphi_k(t) + \epsilon_i(t)$, where $Z_1 \sim N(0, 0.04)$ and $Z_2 \sim N(0, 0.09)$ and $Z_1$ is independent of $Z_2$. 
The intensity function is given by $\lambda\{t \mid \overline{O}_{i}(t)\} =  \exp\{2X_i^{\obs}(t-) - 0.75Z_1 + 0.25Z_2\}$. Consequently, there are 
approximately 12 observations per curve. In this setting, the covariate process $Z_i$ is a time-independent vector $Z_i = (Z_{i1}, Z_{i2})^{\T}$. 

\begin{figure}[H]
	\centering
	{\includegraphics[width=14cm]{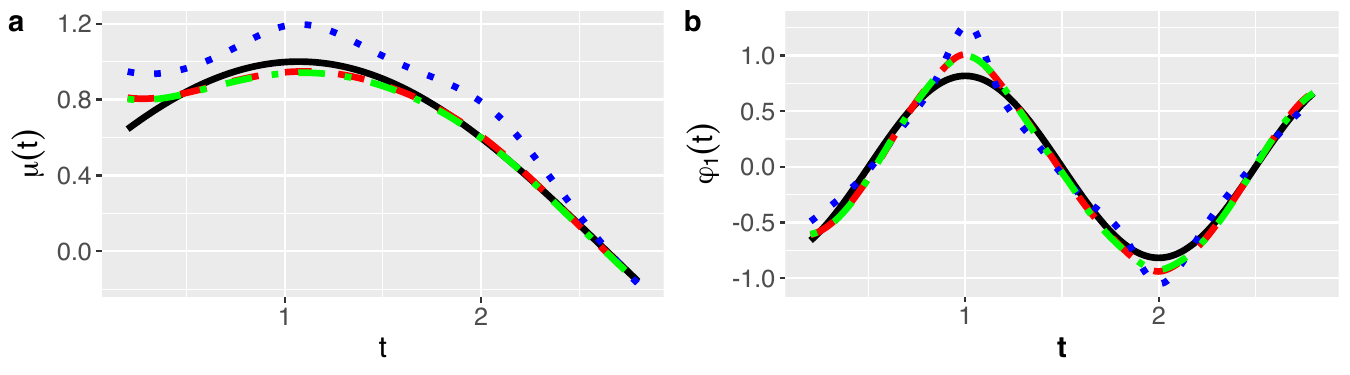}}
	\caption{
		Simulation results of the average of estimated mean function (Panel a) and the average of  estimated first functional principal component (Panel b) across 200 Monte Carlo replicates in Setting 5. In both panels, the black solid line denotes the true function, while the red dashed, green dash-dotted, and blue dotted lines denote the estimates from the proposed method with estimated weights, the proposed method with true weights, and the unweighted method, respectively.}
	\label{fig:FS_S5}
\end{figure}

Table \ref{tab:MISE_setting5} summarizes the results of estimating the mean function, the covariance function, and the first functional principal component from the three methods, and Figure \ref{fig:FS_S5} depicts the average of the estimated mean function and the estimated first functional principal component across 200 simulation replicates. As with previous settings, the proposed method still dominates the unweighted method when the covariate process $Z_i(t)$ is a time-independent vector.

\begin{table}[http]
	\def~{\hphantom{0}}
	\centering
	\addtolength{\tabcolsep}{-0.1pt}  
	\caption{ Mean integrated squared errors ($ \times 0.01$) for the estimated mean function, covariance function, and  first functional principal component in Setting 5. {The actual numerical values are the ones displayed in the table multiply by $0.01$.}
		Standard deviations are presented in the bracket.  UW denotes the unweighted method, TW denotes the proposed method assuming that the true intensity function is known, and EW denotes the proposed method where the intensity function is estimated. 
	} 
	\label{tab:MISE_setting5}
	\begin{tabular}{cccccccccccc}
		\multirow{2}{*}{$n$}&
		\multicolumn{3}{c}{$\widehat{\mu}(t)$} & &
		\multicolumn{3}{c}{$\widehat{C}(s, t)$} & & 
		\multicolumn{3}{c}{$\widehat{\varphi}_1(t)$}\\

		& UW & TW & EW  & & UW & TW & EW  & & UW & TW  & EW \\
		100	& 5.94 &  1.19 &  1.09  &  & 3.57 & 2.73 & 2.56 & & 16.7 & 12.0 & 11.5  \\
		& (2.07) & (.82) & (.67) & & (2.93) & (.98) & (.74) & & (11.0) & (6.5) & (6.5)\\
		
		& & & & & & & & & & &  \\
		200	&  6.14 &  .84 &  .81   &  & 3.06 & 2.06 & 1.96  &  & 15.2 & 8.6 & 8.0   \\
		& (1.47) & (.45) & (.47) & & (2.35) &  (.69) & (.58) & & (9.0) & (4.8) & (3.8) \\

	\end{tabular}
	
\end{table}

\subsection{Setting 6: observational times and longitudinal outcomes are independent} \label{Ssec:ind}
In this setting, the functional response process $\{X_i(t): i=1, \ldots, n\}$ is generated by $X_i(t) = \sin(t + 1/2) + \sum_{k = 1}^{50} \nu_k \zeta_{ik} \varphi_k(t) + \epsilon_i(t)$, where $\nu_k = (-1)^{k+1} (k + 1)^{-1}$, $\zeta_{ik}$'s are independently following a uniform distribution over $[-\sqrt{3}, \sqrt{3}]$ and $\varphi_k(t) = \sqrt{2/3}\cos(k \pi t) $ for $k \geq 1$ and $\epsilon_i(t)$'s are independently normally distributed across both $i$ and $t$, with mean $0$ and variance $0.01$. For each curve, 12 observational time points are randomly generated from the uniform distribution over $[0, 3]$. We investigated the finite-sample performance of our proposed method when the observation time points are independent of the observed outcomes.

\begin{figure}[H]
	\centering
	{\includegraphics[width=14cm]{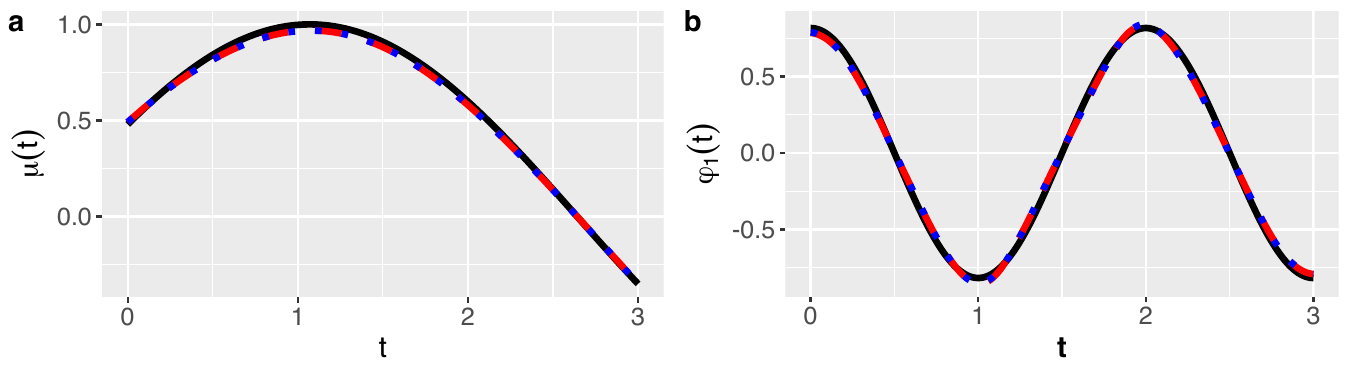}}
	\caption{
		Simulation results of the average of estimated mean function (Panel a) and the average of  estimated first functional principal component (Panel b) across 200 Monte Carlo replicates in Setting 6. In both panels, the black solid line denotes the true function, while the red dashed and blue dotted lines denote the estimates from the proposed method with estimated weights and the unweighted method, respectively.}
	\label{fig:ind}
\end{figure}

Table \ref{tab:MISE_setting6} summarizes the results of estimating the mean function, the covariance function, the first functional principal component, and the first eigenvalue from the unweighted method and the proposed weighted method. Even though the 
observation time points are independent of the observed outcomes, the performance of our proposed weighted method is quite similar to that of the unweighted method. Figure \ref{fig:ind}, which displays the average of the estimated mean function and the estimated first functional principal component across 200 simulation replicates, further justifies this conclusion.

\begin{table}[http]
	\def~{\hphantom{0}}
	\centering
	\addtolength{\tabcolsep}{-1.5pt}  
	\caption{ Mean integrated squared errors ($ \times 0.01$) for the estimated mean function, covariance function, and  first functional principal component in Setting 6. {The actual numerical values are the ones displayed in the table multiplied by $0.01$.}
		Standard deviations are presented in the bracket.  UW denotes the unweighted method, and EW denotes the proposed method where the intensity function is estimated.
	} 
	\label{tab:MISE_setting6}
	\begin{tabular}{cccccccccccccc}
		\multirow{2}{*}{$n$}&
		\multicolumn{2}{c}{$\widehat{\mu}(t)$} & &
		\multicolumn{2}{c}{$\widehat{C}(s, t)$} & & 
		\multicolumn{2}{c}{$\widehat{\varphi}_1(t)$} & & 
		\multicolumn{2}{c}{$\widehat{\kappa}_1$}\\
		
		& UW & EW  & & UW & EW  & & UW & EW & & UW & EW \\
		100	&  .63 & .62  &  &  2.14 & 2.13 & & 3.96 & 3.93  & & .79 & .82  \\
		& (.40) & (.39) & &  (.33) & (.33) & & (2.11) & (2.15) & & 3.22 & 3.23 \\
		& & & & & & & &  & & &  \\
		200	&  .40 & .39   &  &  1.60 & 1.61  &  & 2.26 & 2.23 & & .67 & .67   \\
		& (.22) & (.21) & &  (.18)& (.19) & & (1.08) & (1.11) & & 2.31 & 2.33  \\

		
	\end{tabular}

\end{table}

\section{Additional results of the real application}
\label{Ssec:application}

\begin{figure}[H]
	\centering
	{\includegraphics[width=14cm]{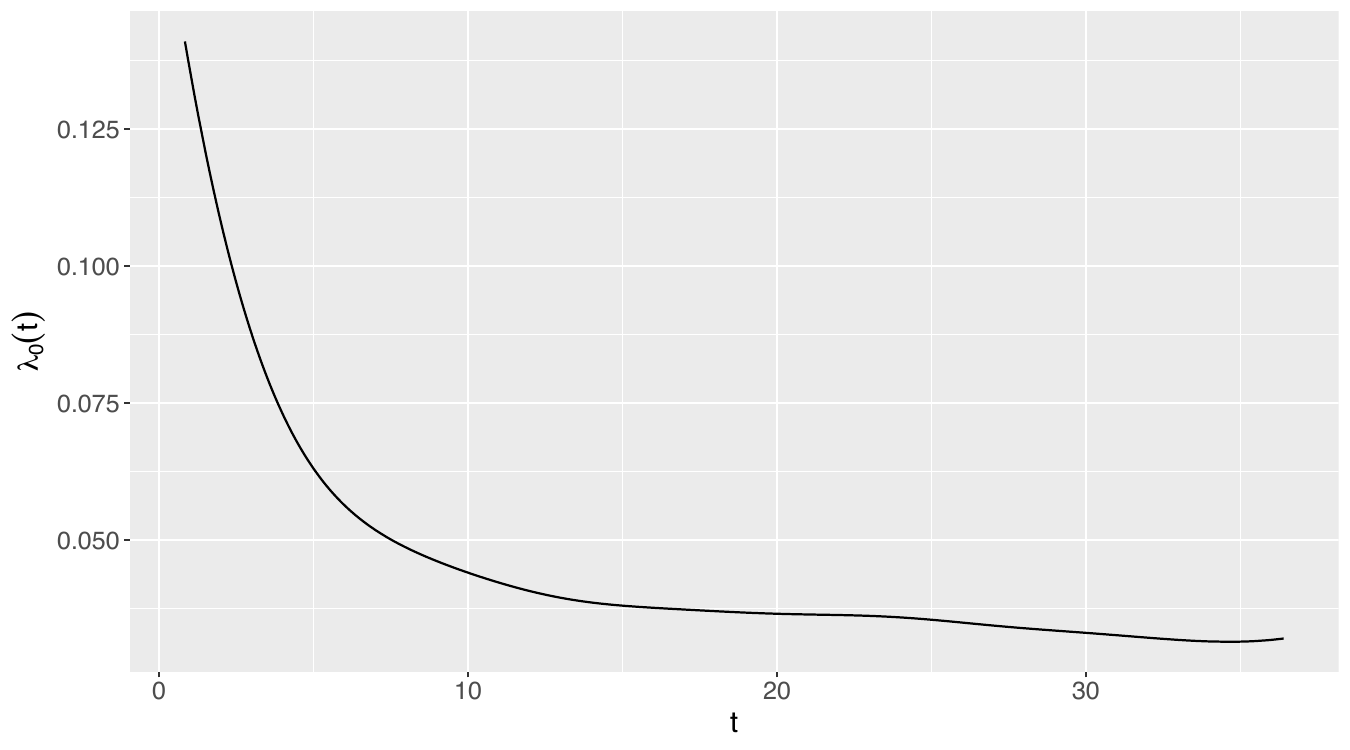}}
	\caption{Estimated baseline intensity for the log CD4 counts.}
	\label{fig:CD4_baseline}
\end{figure}

We model the
intensity of visit times by a Cox proportional intensity function adjusting for log CD4 counts and log viral load at 
the closest past visit.
In the fitted intensity function, the estimated coefficients associated with the log CD4 counts and the log viral load are $\widehat{\beta}_1 = -0.039$ and $\widehat{\beta}_2 = 0.036$, respectively.
The standard errors for $\hat{\beta}_1$ and $\hat{\beta}_2$ are 0.083 and 0.013, respectively. The p-values are 0.66 and 0.01, respectively. So the first coefficient (the log CD4) is not significant while the second (the log viral load) is significant. 
The fitted result shows that patients with lower CD4 counts and higher log viral load are more likely to visit. Figure \ref{fig:CD4_baseline} depicts the estimated baseline intensity function, $\hat{\lambda}_0(t)$. 

\begin{figure}[H]
	\centering
	{\includegraphics[width=14cm]{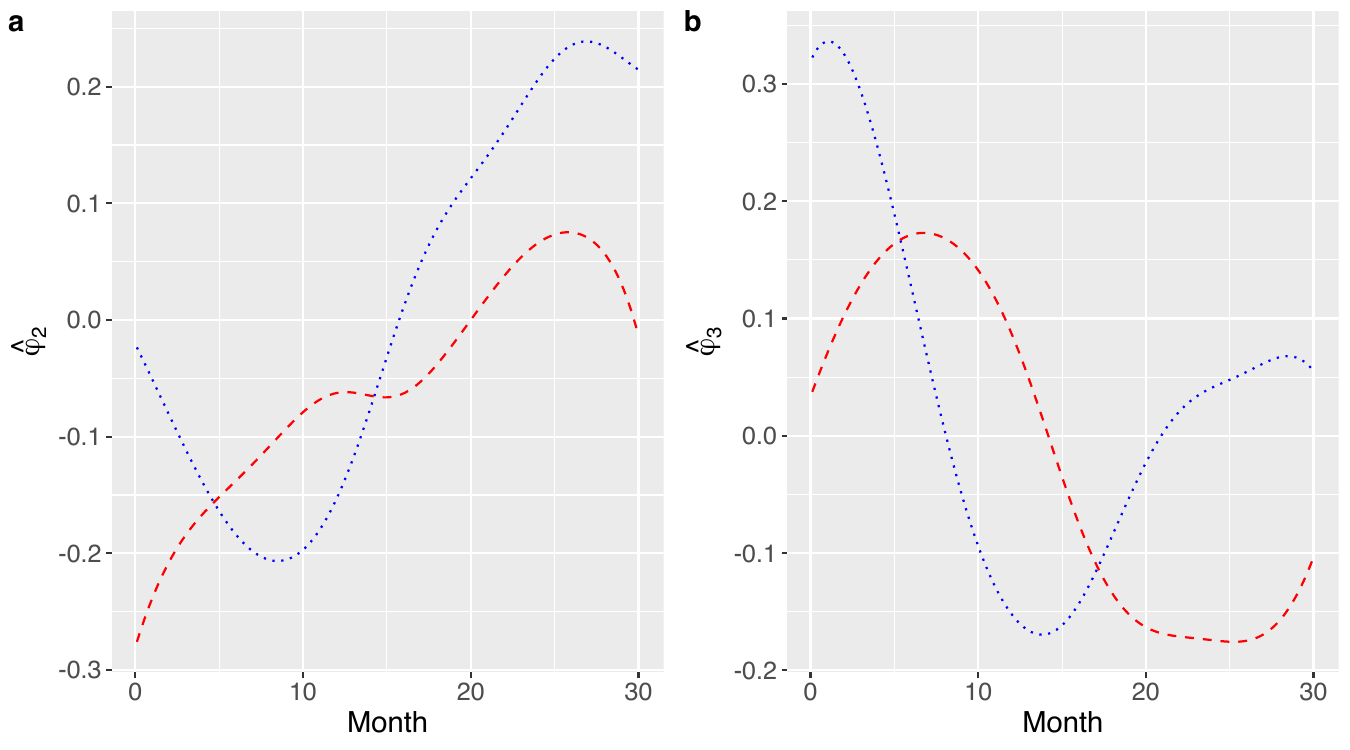}}
	\caption{(a) \& (b) The second and the third estimated eigenfunctions  of {log} CD4 counts from the unweighted method and the proposed weighted method. Here red dashed and blue dotted lines  represent the estimates from the weighted and unweighted methods, respectively.}
	\label{fig:CD4eigf23}
\end{figure}

\begin{figure}[H]
	\centering
	{\includegraphics[width=14cm]{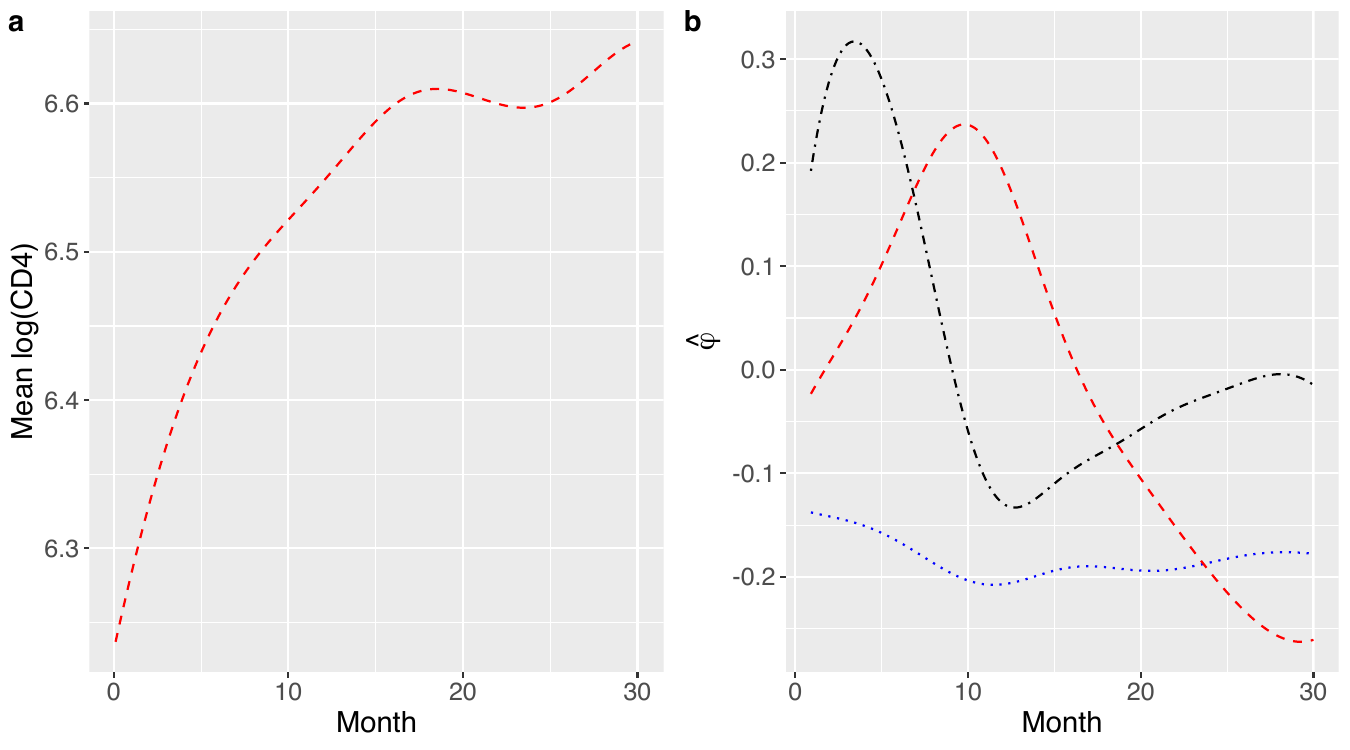}}
	\caption{(a) The estimated mean trajectory of the log CD4 counts from applying the proposed weights to PACE. (b) The first to  the third estimated eigenfunctions  of {log} CD4 counts from applying the proposed weights to PACE. Here red dashed, blue dotted,  and black dotdash lines  represent the estimates of the first, second and third eigenfunctions, respectively.}
	\label{fig:CD4pace}
\end{figure}

Figure \ref{fig:CD4eigf23} displays the second and the third estimated functional principal components from the unweighted method and our proposed method. As suggested by one referee, we apply our proposed weight to the method of Principal Analysis by Conditional Expectation (PACE) proposed by \cite{yao2005}. The left panel of Figure \ref{fig:CD4pace} shows the estimated mean function of the log CD4 counts, while the right panel displays the estimates of the first three functional principal components. The estimates are similar to those generated from our proposed method.

\begin{figure}[H]
	\centering
	{\includegraphics[width=14cm]{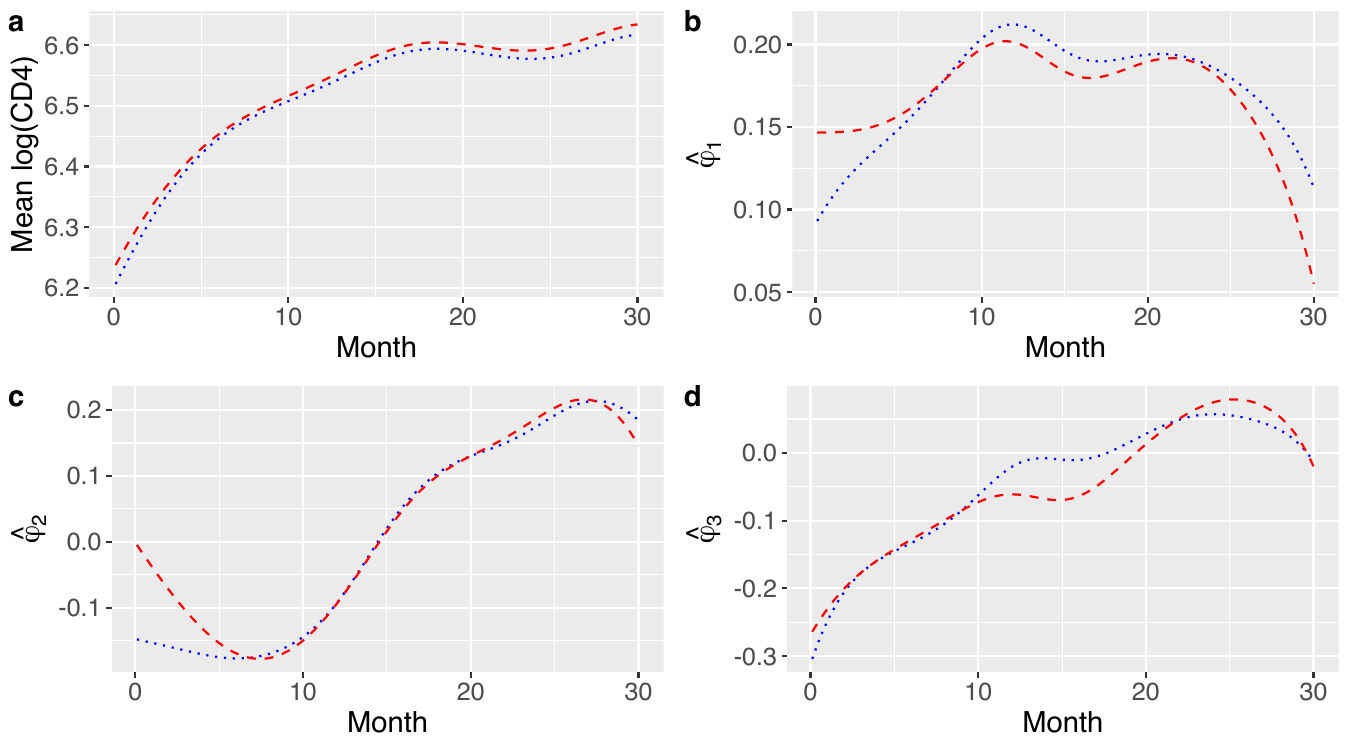}}
	\caption{(a) The estimated mean trajectory of the log CD4 counts from the proposed method with the new intensity function. (b) (c) \& (d) The first to  the third estimated eigenfunctions  of {log} CD4 counts from the proposed method with the new intensity function. Here red dashed and blue dotted lines represent the estimates from the weighted and unweighted methods, respectively. }
	\label{fig:CD4intens2}
\end{figure}

To investigate the effect of the choice of $Z(t)$ in the resulting functional principal component analysis, we conduct a sensitivity analysis. 
\blue{We consider another intensity function, where $g(\overline{O}_i(t))$ is taken as the log viral load and its square.}
Figure \ref{fig:CD4intens2} displays the corresponding results of the estimated mean function and the estimated first three functional principal components for the log CD4 counts. Compared with Figure \ref{fig:CD4} in the main text and Figure \ref{fig:CD4eigf23}, we find that such choice of $g(\overline{O}_i(t))$ indeed leads to a new intensity function, but does not lead to an obvious change in these estimates. 

\begin{figure}[ht]
	\centering
	{\includegraphics[width=14cm]{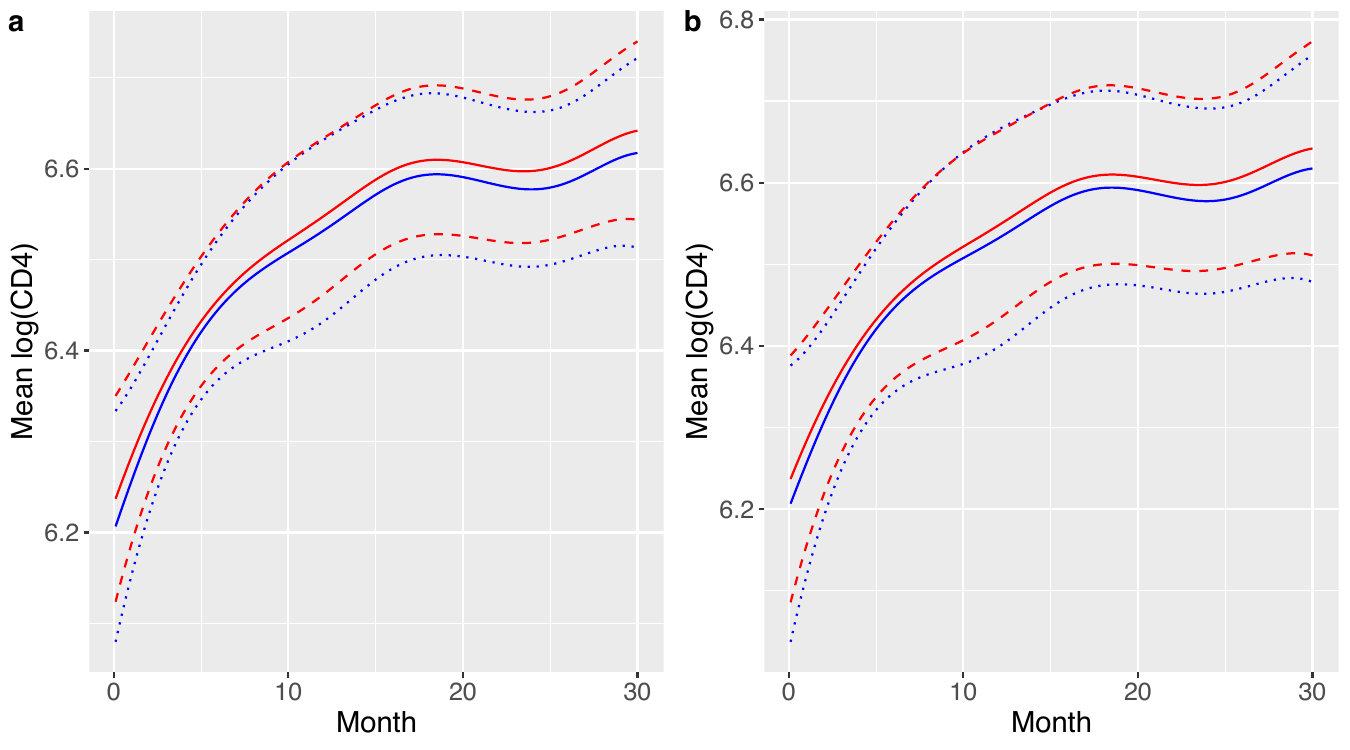}}
	\caption{(a) The bootstrap pointwise confidence intervals for the mean function based on these two methods. (b) The bootstrap simultaneous confidence bands for the mean function based on these two methods. Here red and blue lines represent the estimates from the weighted and unweighted methods, respectively.}
	\label{fig:CD4meancb}
\end{figure}

\begin{figure}[ht]
	\centering
	{\includegraphics[width=14cm]{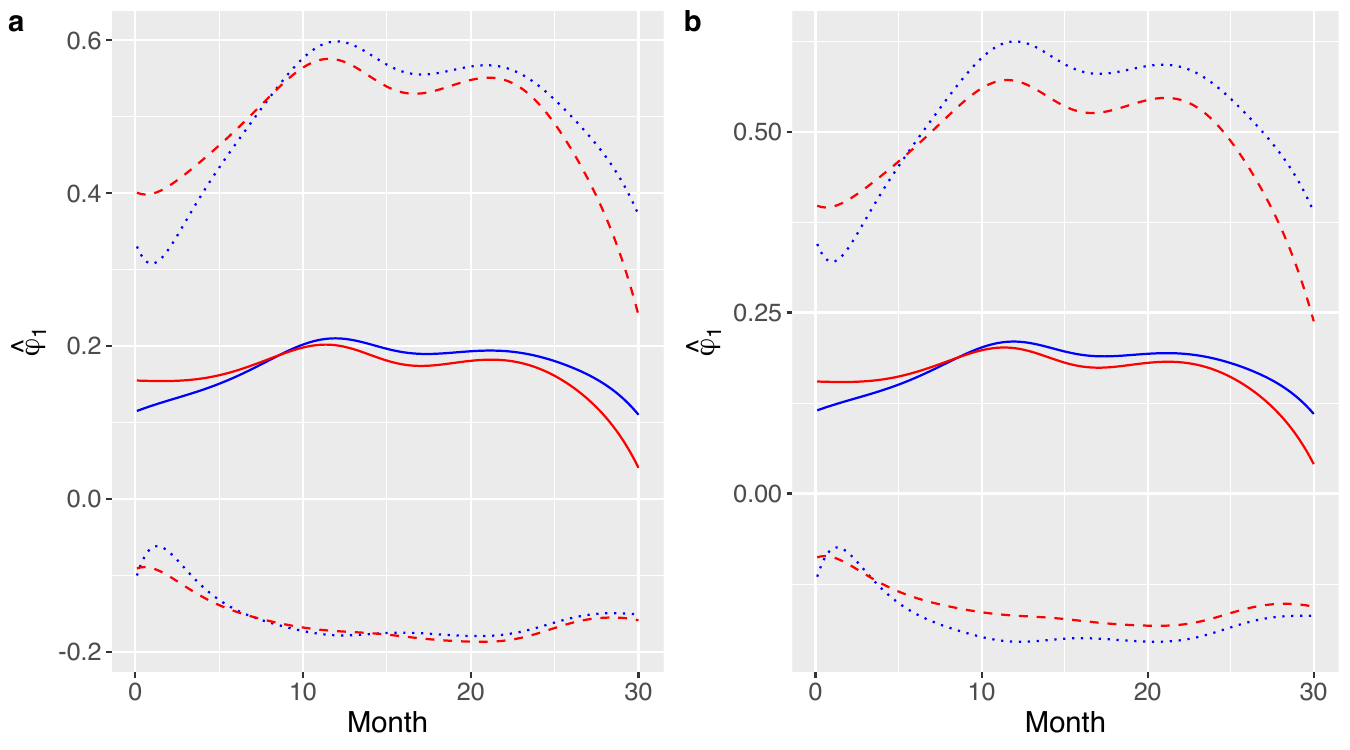}}
	\caption{(a) The bootstrap pointwise confidence intervals for the first eigenfunction based on these two methods. (b) The bootstrap simultaneous confidence bands for the first eigenfunction based on these two methods. Here red and blue lines represent the estimates from the weighted and unweighted methods, respectively.}
	\label{fig:CD4PCcb}
\end{figure}

To assess the difference in the estimated mean functions between these methods, we employ the bootstrap method to generate pointwise confidence intervals and simultaneous confidence bands. Figure \ref{fig:CD4meancb} depicts the bootstrap pointwise and simultaneous confidence bands for the mean function based on these two methods. It suggests there does exist significant difference between these two estimates for this application. Moreover, Figure \ref{fig:CD4PCcb} displays the bootstrap pointwise and simultaneous confidence bands for the first eigenfunction of the covariance function based on these two methods. We still cannot find a noticeable difference between these two estimates for this application. 

\bibliographystyle{natbib}
\bibliography{Ref}

\end{document}